\documentclass[longauth]{aa}

\usepackage{float}
\usepackage{txfonts}
\usepackage{graphicx}
\usepackage{natbib}
\usepackage{xspace}
\bibpunct{(}{)}{;}{a}{}{,}

\def\mhmpc{\,h^{-1}{\rm Mpc}}

\def\upstrut{\phantom{\rlap{${U}^{U^U}$}}}
\def\downstrut{\phantom{\rlap{${L}_{L_L}$}}}

\newcommand{\hmpc}{$\,h^{-1}$ Mpc\xspace}
\newcommand{\kms}{$\,{\rm km\, s^{-1}}$\xspace}
\newcommand{\SSR}{${SSR}$\xspace}
\newcommand{\TSR}{${TSR}$\xspace}
\newcommand{\CSR}{${CSR}$\xspace}

\newcommand{\bs}[1]{\boldsymbol{#1}}

\def\citejap#1{\citeauthor{#1}\ \citeyear{#1}}
\def\CSR{{\rm CSR}}
\def\SSR{{\rm SSR}}
\def\TSR{{\rm TSR}}

\begin{document}

\title{The VIMOS Public Extragalactic Redshift Survey (VIPERS)\thanks{
    Based on observations collected at the European Southern
    Observatory, Cerro Paranal, Chile, using the Very Large Telescope
    under programs 182.A-0886 and partly 070.A-9007.  Also based on
    observations obtained with MegaPrime/MegaCam, a joint project of
    CFHT and CEA/DAPNIA, at the Canada-France-Hawaii Telescope (CFHT),
    which is operated by the National Research Council (NRC) of
    Canada, the Institut National des Sciences de l’Univers of the
    Centre National de la Recherche Scientifique (CNRS) of France, and
    the University of Hawaii. This work is based in part on data
    products produced at TERAPIX and the Canadian Astronomy Data
    Centre as part of the Canada-France-Hawaii Telescope Legacy
    Survey, a collaborative project of NRC and CNRS. The VIPERS web
    site is http://www.vipers.inaf.it/.}  }

\subtitle{The growth of structure at $\mathbf{0.5<z<1.2}$ from
  redshift-space distortions in the clustering of the PDR-2 final
  sample}

\titlerunning{Redshift-space distortions in VIPERS}

\author{
A.~Pezzotta\inst{\ref{brera},\ref{bicocca}}
\and S.~de la Torre\inst{\ref{lam}}       
\and J.~Bel\inst{\ref{cpt},\ref{brera}}
\and B.~R.~Granett\inst{\ref{brera},\ref{unimi}}                                                 
\and L.~Guzzo\inst{\ref{brera},\ref{unimi}}      
\and J.~A.~Peacock\inst{\ref{roe}}
%
%
\and B.~Garilli\inst{\ref{iasf-mi}}          
\and M.~Scodeggio\inst{\ref{iasf-mi}}       
\and M.~Bolzonella\inst{\ref{oabo}}      
%
%
\and U.~Abbas\inst{\ref{oa-to}}
\and C.~Adami\inst{\ref{lam}}
\and D.~Bottini\inst{\ref{iasf-mi}}
\and A.~Cappi\inst{\ref{oabo},\ref{nice}}
\and O.~Cucciati\inst{\ref{unibo},\ref{oabo}}           
\and I.~Davidzon\inst{\ref{lam},\ref{oabo}}   
\and P.~Franzetti\inst{\ref{iasf-mi}}   
\and A.~Fritz\inst{\ref{iasf-mi}}       
\and A.~Iovino\inst{\ref{brera}}
\and J.~Krywult\inst{\ref{kielce}}
\and V.~Le Brun\inst{\ref{lam}}
\and O.~Le F\`evre\inst{\ref{lam}}
\and D.~Maccagni\inst{\ref{iasf-mi}}
\and K.~Ma{\l}ek\inst{\ref{warsaw-nucl}}
\and F.~Marulli\inst{\ref{unibo},\ref{infn-bo},\ref{oabo}} 
\and M.~Polletta\inst{\ref{iasf-mi},\ref{marseille-uni},\ref{toulouse}}
\and A.~Pollo\inst{\ref{warsaw-nucl},\ref{krakow}}
\and L.A.M.~Tasca\inst{\ref{lam}}
\and R.~Tojeiro\inst{\ref{st-andrews}}  
\and D.~Vergani\inst{\ref{iasf-bo}}
\and A.~Zanichelli\inst{\ref{ira-bo}}
%
%
\and S.~Arnouts\inst{\ref{lam},\ref{cfht}} 
\and E.~Branchini\inst{\ref{roma3},\ref{infn-roma3},\ref{oa-roma}}
\and J.~Coupon\inst{\ref{geneva}}
\and G.~De Lucia\inst{\ref{oats}}
\and J.~Koda\inst{\ref{brera}}
\and O.~Ilbert\inst{\ref{lam}}
\and F. G.~Mohammad\inst{\ref{brera}}
\and T.~Moutard\inst{\ref{halifax},\ref{lam}}  
%
%
\and L.~Moscardini\inst{\ref{unibo},\ref{infn-bo},\ref{oabo}}
%
}

\institute{
INAF - Osservatorio Astronomico di Brera, Via Brera 28, 20122 Milano
--  via E. Bianchi 46, 23807 Merate, Italy \label{brera}
\and Dipartimento di Fisica, Universit\`a di Milano-Bicocca, P.zza della Scienza 3, I-20126 Milano, Italy \label{bicocca}
\and Aix Marseille Univ, CNRS, LAM, Laboratoire d'Astrophysique de
Marseille, Marseille, France  \label{lam}
\and Aix Marseille Univ, Univ Toulon, CNRS, CPT, Marseille, France \label{cpt}
\and  Universit\`{a} degli Studi di Milano, via G. Celoria 16, 20133 Milano, Italy \label{unimi}
\and Institute for Astronomy, University of Edinburgh, Royal
Observatory, Blackford Hill, Edinburgh EH9 3HJ, UK \label{roe}
\and INAF - Istituto di Astrofisica Spaziale e Fisica Cosmica Milano, via Bassini 15, 20133 Milano, Italy \label{iasf-mi}
\and INAF - Osservatorio Astronomico di Bologna, via Ranzani 1, I-40127, Bologna, Italy \label{oabo} 
\and INAF - Osservatorio Astrofisico di Torino, 10025 Pino Torinese, Italy \label{oa-to}
\and Laboratoire Lagrange, UMR7293, Universit\'e de Nice Sophia Antipolis, CNRS, Observatoire de la C\^ote d’Azur, 06300 Nice, France \label{nice}
\and Dipartimento di Fisica e Astronomia - Alma Mater Studiorum Universit\`{a} di Bologna, viale Berti Pichat 6/2, I-40127 Bologna, Italy \label{unibo}
\and Institute of Physics, Jan Kochanowski University, ul. Swietokrzyska 15, 25-406 Kielce, Poland \label{kielce}
\and National Centre for Nuclear Research, ul. Hoza 69, 00-681 Warszawa, Poland \label{warsaw-nucl}
\and INFN, Sezione di Bologna, viale Berti Pichat 6/2, I-40127 Bologna, Italy \label{infn-bo}
\and Aix-Marseille Université, Jardin du Pharo, 58 bd Charles Livon, F-13284 Marseille cedex 7, France \label{marseille-uni}
\and IRAP,  9 av. du colonel Roche, BP 44346, F-31028 Toulouse cedex 4, France \label{toulouse} 
\and Astronomical Observatory of the Jagiellonian University, Orla 171, 30-001 Cracow, Poland \label{krakow} 
\and School of Physics and Astronomy, University of St Andrews, St Andrews KY16 9SS, UK \label{st-andrews}
\and INAF - Istituto di Astrofisica Spaziale e Fisica Cosmica Bologna, via Gobetti 101, I-40129 Bologna, Italy \label{iasf-bo}
\and INAF - Istituto di Radioastronomia, via Gobetti 101, I-40129,Bologna, Italy \label{ira-bo}
\and Canada-France-Hawaii Telescope, 65--1238 Mamalahoa Highway, Kamuela, HI 96743, USA \label{cfht}
\and Dipartimento di Matematica e Fisica, Universit\`{a} degli Studi Roma Tre, via della Vasca Navale 84, 00146 Roma, Italy\label{roma3} 
\and INFN, Sezione di Roma Tre, via della Vasca Navale 84, I-00146 Roma, Italy \label{infn-roma3}
\and INAF - Osservatorio Astronomico di Roma, via Frascati 33, I-00040 Monte Porzio Catone (RM), Italy \label{oa-roma}
\and Department of Astronomy, University of Geneva, ch. d’Ecogia 16, 1290 Versoix, Switzerland \label{geneva}
\and INAF - Osservatorio Astronomico di Trieste, via G. B. Tiepolo 11, 34143 Trieste, Italy \label{oats}
\and Department of Astronomy \& Physics, Saint Mary's University, 923 Robie Street, Halifax, Nova Scotia, B3H 3C3, Canada \label{halifax}
%
%
}

\authorrunning{A. Pezzotta et al.}

\offprints{\mbox{A.~Pezzotta},\\ \email{andrea.pezzotta@brera.inaf.it}}

\abstract{We present measurements of the growth rate of cosmological
  structure from the modelling of the anisotropic galaxy clustering
  measured in the final data release of the VIPERS survey. The
  analysis is carried out in configuration space and based on
  measurements of the first two even multipole moments of the
  anisotropic galaxy auto-correlation function, in two redshift bins
  spanning the range $0.5 < z < 1.2$. We provide robust and
  cosmology-independent corrections for the VIPERS angular selection
  function, allowing recovery of the underlying clustering amplitude
  at the percent level down to the $\mathrm{Mpc}$ scale. We discuss
  several improvements on the non-linear modelling of redshift-space
  distortions (RSD) and perform detailed tests of a variety of
  approaches against a set of realistic VIPERS-like mock
  realisations. This includes using novel fitting functions to
  describe the velocity divergence and density power spectra
  $P_{\theta\theta}$ and $P_{\delta\theta}$ that appear in RSD models.
  These tests show that we are able to measure the growth rate with
  negligible bias down to separations of $5\mhmpc$.  Interestingly,
  the application to real data shows a weaker sensitivity to the
  details of non-linear RSD corrections compared to mock results. We
  obtain consistent values for the growth rate times the matter power
  spectrum normalisation parameter of $f\sigma_8=0.55\pm 0.12$ and
  $0.40\pm0.11$ at effective redshifts of $z = 0.6$ and $z=0.86$
  respectively. These results are in agreement with standard cosmology
  predictions assuming Einstein gravity in a $\Lambda \rm{CDM}$
  background.}


\maketitle

\section{Introduction}

The discovery of the accelerated expansion of the Universe in the late
stages of the 20th Century has given us a self-consistent standard
cosmological model, which is in close agreement with virtually all
current cosmological observations. Multiple lines of evidence, from
cosmic microwave background anisotropies \citep{hinshaw12,planck15},
baryon acoustic oscillations in the galaxy distribution
\citep{beutler11,blake11,anderson12}, to SNe Ia luminosity distances
\citep{riess98,perlmutter99}, require most of the energy content of
the Universe to be in form of a repulsive `dark energy' that is
empirically close in behaviour to the classical cosmological constant
(see e.g. \citejap{weinberg13} for some history and a review of
current constraints).  The nature of dark energy is naturally a
question of huge interest, with possibilities ranging from a fixed
vacuum energy density with equation of state $w=P/\rho c^2=-1$, to
dynamical models based on evolving scalar fields varying both in space
and time. Such models motivate an effort to measure $w$ and its
evolution. But independently of the outcome of this exercise, it
remains the puzzle that a very large vacuum density seems to be
necessary -- so the much smaller observed value therefore requires a
challenging degree of fine tuning \citep{weinberg89}.

A more radical explanation for the observed acceleration could be that
the theory of gravity itself is modified on cosmological scales
\citep{carroll04,jain2010,clifton11}. Commonly discussed alternatives
include $f(R)$ gravity, where the gravitational Lagrangian is made
more complicated than a simple Ricci scalar $R$; chameleon models that
invoke a fifth fundamental force to drive the acceleration; and DGP
(Dvali-Gabadadze-Porrati) models, which postulate a higher-dimensional
Minkowski space-time, within which the ordinary 3+1 space-time is
embedded.  For an appropriate choice of model parameters, dark energy
and modified gravity can both reproduce the observed expansion history
$H(z)$. In principle this degeneracy can be lifted by measuring the
growth rate of cosmic structure. Modifications of gravity involve a
variation in the strength of the gravitational force with scale or
environment, and thus a key question is whether density fluctuations
are growing at the rate predicted by models involving General
Relativity and a homogeneous dark energy.

Among observational methods to estimate the growth rate of structure,
{\sl redshift-space distortions} (RSD) in the clustering pattern of
galaxies \citep{kaiser87} have assumed a growing importance in the
last decade \citep[e.g.][]{guzzo08}. RSD arise when the Doppler effect
of galaxy peculiar velocities supplements the isotropic Hubble
expansion. Peculiar velocities are inevitably associated with
gravitational growth of inhomogeneities, which can be described by the
logarithmic growth rate of density perturbations:
\begin{equation}
f\equiv\frac{d\ln\delta}{d\ln a}\,\,\,\,\, ,
\end{equation}
where $\delta$ is the fractional density fluctuation, and $a$ is the
cosmic scale factor. For many (but not all) theories of gravity, this
growth rate can be well approximated by an empirical relation as
$f(z)=[\Omega(z)]^\gamma$ \citep{peebles80,lahav1991}, provided the
fluctuations are in the linear regime and in the growing mode. For
Einstein gravity, $\gamma\simeq 0.55$; but this parameter can vary by
around 0.1 between different commonly-discussed models of late-time
dark energy and modified gravity
\citep{dvali00,linder2007}. Measurements of linear RSD from galaxy
redshift surveys constrain the combination $\beta=f/b$, where $b$ is
an unknown linear galaxy bias parameter.  But the real-space galaxy
autocorrelation function, $b^2\xi_{\rm mass}$, is observable -- so
this can be eliminated to yield an estimate of a quantity that purely
concerns dark matter: $f\sigma_8$, with $\sigma_8$ being the {\it rms}
linear matter fluctuations within spheres of radius $8\mhmpc$.

\begin{figure*}
\centering
\includegraphics[width=\textwidth]{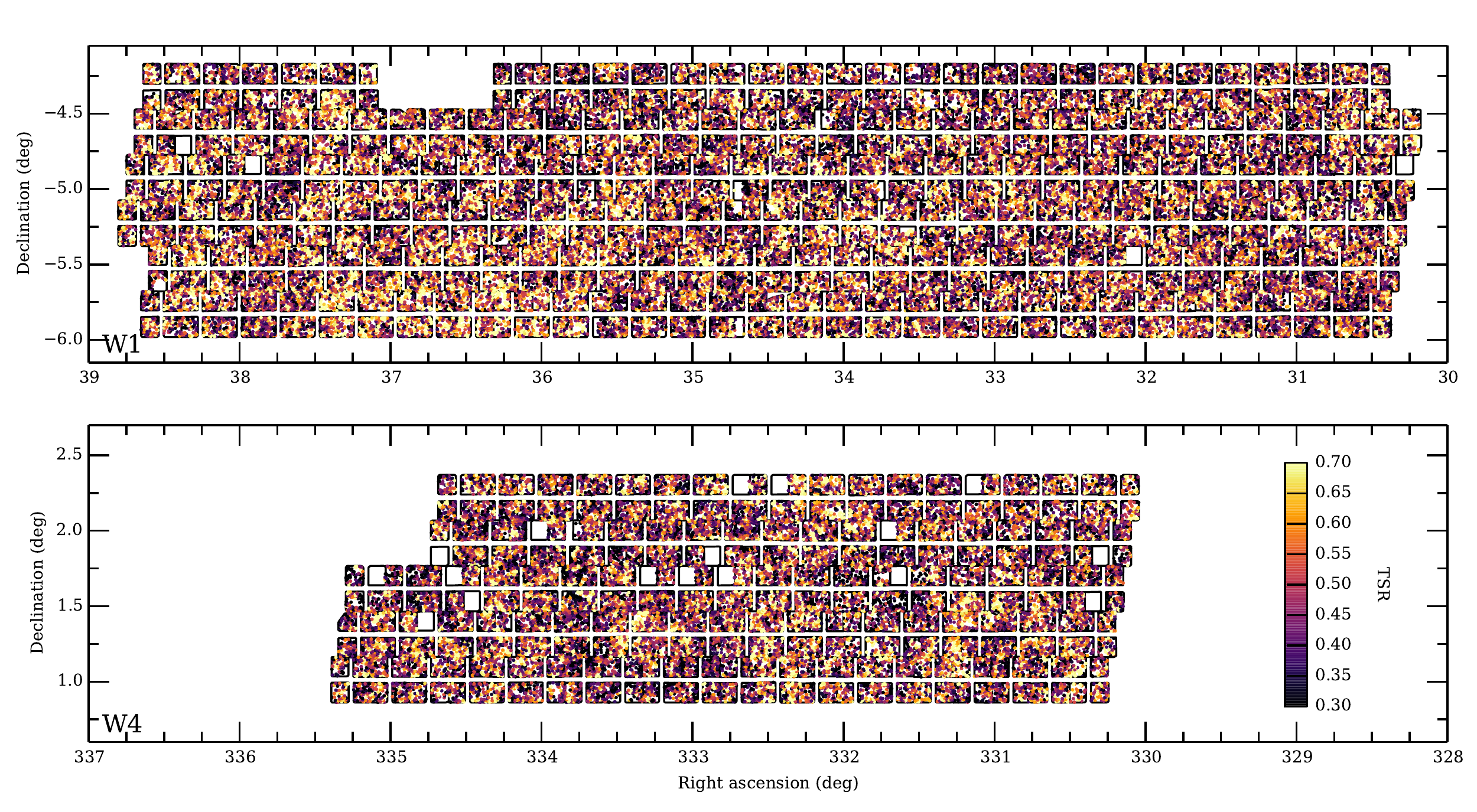}
\caption{Footprint of the VIPERS observations within the W1 (top) and
  W4 (bottom) fields, as reconstructed from the final galaxy
  sample. The VIMOS pointings and quadrants are marked by black
  rectangles. Galaxies are colour-coded according to their value of
  the Target Sampling Rate (TSR: see Sect.
  \ref{sec:angular_effects}), which can be considered as a proxy for
  the inverse of the projected galaxy density field. Empty rectangles
  correspond to failed quadrants, for which the spectroscopic mask
  insertion failed or was incorrect, leading to no data being
  collected.  }
\label{fig:vipers_field}
\end{figure*}

Unfortunately, extracting the linear RSD signal from galaxy redshift
surveys is non-trivial, because much of the RSD signal lies on
quasi-linear and non-linear scales.  A simple and widely-used
extension of the linear Kaiser model is the `dispersion model'
\citep{peacock94}, which accounts for radial convolution by a random
velocity dispersion plus non-linear
corrections to the real-space power spectrum.  This model was
successfully applied to several galaxy surveys in the past
\citep{peacock01,guzzo08}, but is insufficiently accurate to be
trusted when the precision allowed by the data goes below 10\%
(\citejap{okumura11}, \citejap{bianchi12}; see also the companion
paper by \citejap{wilson16}).  There have been a number of attempts to
derive improved RSD models.  As shown by \citet{scoccimarro04}, the
dispersion model is a simplification of the original {\sl streaming
  model} \citep{peebles80,fisher95}, in which the full redshift-space
correlation function is obtained by convolution with a proper
scale-dependent pairwise velocity distribution.  But predicting this
distribution function is hard \cite[e.g.][]{bianchi15a,uhlemann15},
and typical applications simplify the problem by adopting a
(scale-dependent) Gaussian pairwise distribution function
\citep[e.g.][]{reid12}.  \citet{scoccimarro04} proposed an influential
alternative, in which the linear Kaiser term is generalised by
including the velocity and velocity-density power spectra. This
concept was extended by the TNS model \citep{taruya10}, which takes
better into account the non-linear coupling between the density and
the velocity field. This model is currently considered as one of the
best descriptions of RSD down to the quasi-linear regime.

These theoretical developments have been stimulated by a growing
number of new measurements from larger datasets. These included in
particular the 6dfGS \citep{beutler12}, WiggleZ
\citep[e.g.][]{blake11,contreras13} and BOSS
\citep[e.g.][]{reid14,beutler16,satpathy16,sanchez16,grieb16}.  The
present paper is one in a series aimed at extending this RSD work to
higher redshifts by analysing the final PDR-2 release of the VIMOS
Public Extragalactic Redshift Survey
\cite[VIPERS,][]{guzzo14,garilli14,scodeggio16}.  This survey has
collected redshifts for about $90\,000$ galaxies in the range $0.4
\lesssim z \lesssim 1.2$ with sampling and volume comparable to those
of local surveys, such as the Two-degree Field Galaxy Redshift Survey
(2dFGRS) at $z\simeq0.1$ \citep{colless01}.  The prime original goal
of VIPERS was an accurate measurement of the growth rate of structure
at redshift around unity. An early measurement was performed using the
Public Data Release 1 (PDR-1: \citejap{garilli14}), setting a
reference measurement of $f\sigma_8$ at $z = 0.8$
\citep{delatorre13}. Having nearly doubled the sample, this analysis
is now revisited, and expanded in a number of ways.
\citet{delatorre16} performs a configuration space joint analysis
involving RSD and galaxy-galaxy lensing, while \citet{wilson16}
develops a direct Fourier-space approach coupled with the so-called
`clipping' linearisation of the density field; with a similar aim,
\citet{mohammad17} identifies optimal sub-classes of RSD tracers,
focusing on luminous blue galaxies; the analysis presented here uses
the configuration-space information contained in the first two even
multipole moments of the anisotropic correlation function,
implementing the currently most advanced non-linear corrections and
testing their performances on VIPERS-like mocks.

The paper is organised as follows. In Sect. 2 we give a description of
the final VIPERS dataset and of the corresponding mock catalogues used
throughout the analysis, while in Sect. 3 we describe the estimation of
the two-point correlation function of galaxies in redshift
space. Section 4 describes the target selection biases and how these are
mitigated. In Sect. 5 we present the VIPERS measurements.  The error
estimates are described in Sect. 6 along with the fitting
procedure. Section 7 gives a description of the RSD models that are used
in Sect. 8 to understand the level of systematics in the recovery of
the growth rate of structure. The results are presented in Sect. 9 and
discussed in Sect. 10 with our conclusions.

Throughout this analysis, if not specified otherwise, we assume a
fiducial flat $\Lambda{\rm CDM}$ cosmological model with
$(\Omega_m,\Omega_b,n_s)=(0.30,0.045,0.96)$ and parametrise the Hubble
constant as $H_0=100\,h\,\rm{km\,s^{-1}Mpc^{-1}}$.

\section{The VIPERS survey}

\subsection{Observations}
\label{sec:data}

The VIPERS survey covers an overall area of $ 23.5$ deg$^2$ over the
W1 and W4 fields of the Canada-France-Hawaii Telescope Legacy Survey
Wide (CFHTLS-Wide).  The VIMOS multi-object spectrograph
\citep{lefevre03} was used to cover these two fields with a mosaic of
288 pointings, 192 in W1 and 96 in W4 (see Fig.
\ref{fig:vipers_field}). Galaxies are selected from the CFHTLS
catalogue to a faint limit of $i_{\rm AB}=22.5$, applying an
additional $(r-i)$ vs $(u-g)$ colour pre-selection that efficiently
and robustly removes galaxies at $z<0.5$. Coupled with a highly
optimised observing strategy \citep{scodeggio09}, this doubles the
mean galaxy sampling efficiency in the redshift range of interest,
compared to a purely magnitude-limited sample, bringing it to 47\%.

Spectra are collected at moderate resolution ($R\simeq 220$) using the
LR Red grism, providing a wavelength coverage of
5500-9500$\smash{\rm{\AA}}$. The typical redshift error for the sample
of reliable redshifts is $\sigma_z=0.00054(1+z)$, which corresponds to
an error on a galaxy peculiar velocity at any redshift of 163~\kms.
These and other details are given in the full PDR-2 release
accompanying paper \citep{scodeggio16}. A discussion of the data
reduction and management infrastructure was presented in
\citet{garilli14}, while a complete description of the survey design
and target selection was given in the survey description paper
\citep{guzzo14}.  The dataset used in this paper is an early version
of the PDR-2 data, from which it differs by a few hundred redshifts
revised during the very last period before the release. In total it
includes $89\,022$ objects with measured redshifts. As in all
statistical analyses of the VIPERS data, only measurements with
quality flags 2 to 9 inclusive are used, corresponding to a sample with a
redshift confirmation rate of $96.1\%$ (for a description of the
quality flag scheme, see \citejap{scodeggio16}). 

In the analysis presented here we shall analyse two redshift sub-samples of the whole survey (W1 + W4) in the ranges 
$0.5<z<0.7$ and $0.7<z<1.2$, including respectively 30\,764 and 35\,734 galaxies. 

\begin{figure}[t!]
\resizebox{\hsize}{!}{\includegraphics{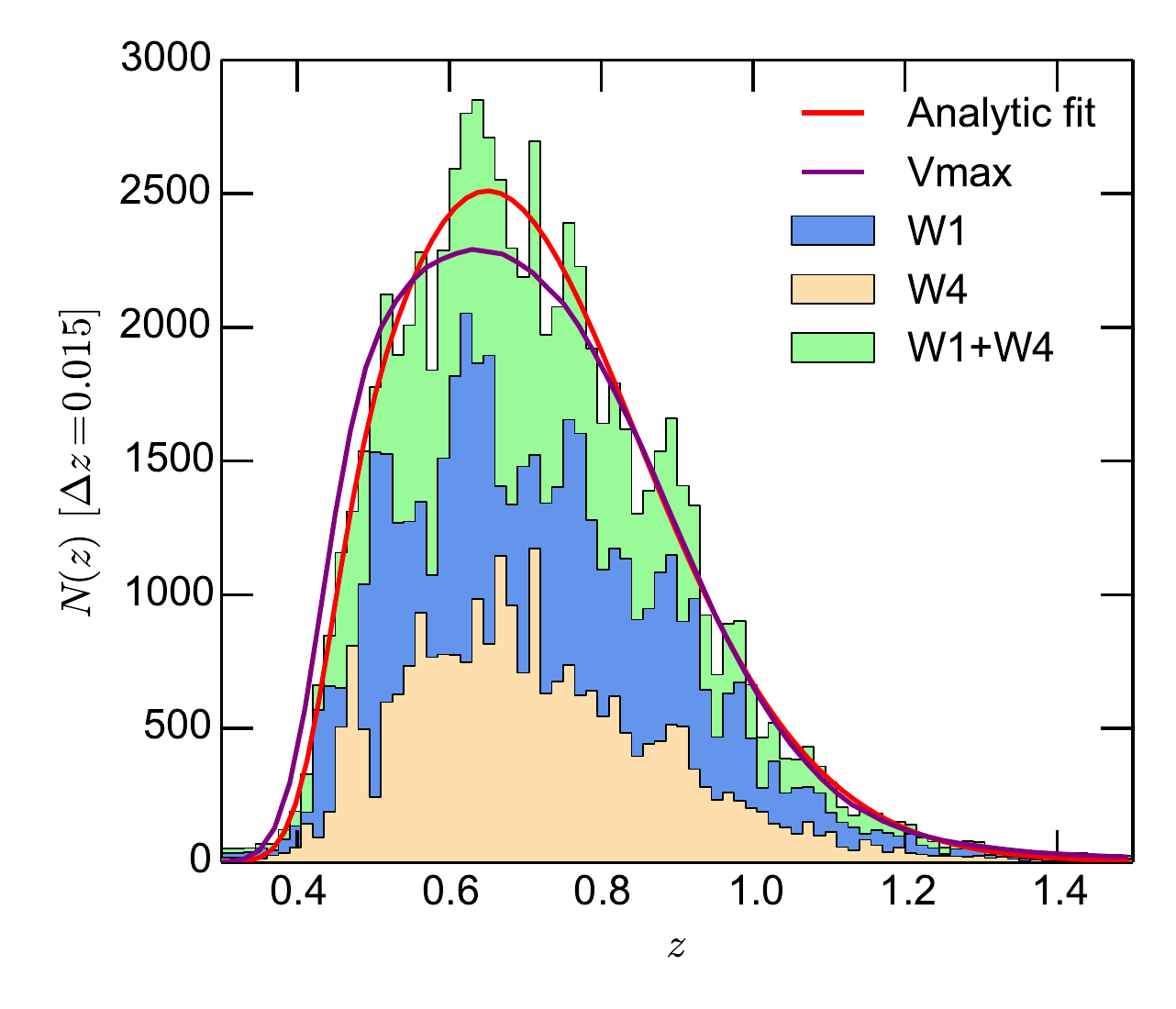}}
\caption{Redshift distribution of the final VIPERS galaxy sample. The
  distributions of redshifts collected separately within the two
  CFHTLS fields are plotted together with the combined distribution
  using different colours. The red and purple solid lines show
  respectively the best fit using the analytic template in
  Eq.~(\ref{eq:nz}) and the predicted $V_{\rm max}$ profile of
  the combined redshift distribution. The peculiar distribution of the
  VIPERS galaxy sample differs from the typical expectation from a
  magnitude-limited sample. This deviation is the result of the
  colour-colour pre-selection adopted to reject most galaxies located
  at $z<0.5$.}
\label{fig:redshift_distribution}
\end{figure}

\subsection{Redshift distribution}

The redshift distribution of the galaxy sample is shown in Fig.
\ref{fig:redshift_distribution}.  At $z>0.6$, it follows the typical
decay in the number of objects expected for a magnitude-limited
survey, while the rapid fall of the counts at $z<0.5$ is the result of
the colour-colour pre-selection.  In \citet{delatorre13} it was shown
that this histogram can be modelled analytically by the functional
form
\begin{equation}
N(z)=A\bigg(\frac{z}{z_0}\bigg)^\alpha\exp\bigg(-\bigg(\frac{z}{z_0}\bigg)^\beta\bigg)\,\CSR(z)\,\, ,
\label{eq:nz}
\end{equation}
where $A$, $z_0$, $\alpha$ and $\beta$ are fitting parameters. The
term $\CSR(z)$ (Colour Sampling Rate) describes the colour-colour
pre-selection in terms of an error function transitioning between $0$
and $1$ around redshift $z=0.5$, i.e.
$\CSR(z)=\left(1-\mathrm{erf}\big[b(z_t-z)\big]\right)\,/\,2 $ where
the transition redshift $z_t$ and the transition width $b$ are free
parameters. As shown in \citet{scodeggio16}, $\CSR(z)$ is unity for
$z\ge 0.6$, corresponding to a purely magnitude-limited selection.

The best fit of Eq.~\ref{eq:nz} to the final VIPERS data is shown
by the red curve in Fig. \ref{fig:redshift_distribution}. Such
modelling of the redshift distribution is an important and sensitive
ingredient when estimating galaxy clustering, as we discuss in
Sect.~\ref{sec:estimation} and in \citet{delatorre13}.  We compare it
with the $V_{\rm max}$ technique \citep[e.g.][]{cole11,delatorre13}
shown in Fig. \ref{fig:redshift_distribution} with the purple
curve. Although we find no significant difference in the resulting
clustering between the two methods, here we chose to use the $V_{\rm
  max}$ method, as in the companion paper of \citet{delatorre16}. A
further method often used in the literature is that of smoothing the
observed redshift distribution with a Gaussian kernel (as for instance
in the parallel papers by \citejap{rota16} and \citejap{wilson16}).

\subsection{Mock galaxy samples} \label{sec:mocks}

In order to test the details of the analysis as well as the modelling
of RSD, we make use of a suite of mock galaxy catalogues designed to
be a realistic match to the VIPERS final dataset. These have been
constructed from the Big MultiDark N-body simulation \citep{klypin16},
which assumes a flat $\Lambda$CDM cosmology with $(\Omega_m,
\Omega_\Lambda, \Omega_b, h, n_s, \sigma_8)=(0.307, 0.693, 0.0482,
0.678, 0.960, 0.823)$ and covers a volume of
$15.625\,h^{-3}\,\mathrm{Gpc}^3$. The construction of the mock samples
is described in \citet{delatorre16} and is based on the method
detailed in \citet{delatorre13}. We refer the reader to these papers
for details and only give a brief overview of the adopted method in
the following.

153 independent lightcones have been extracted from the simulation
volume, which follow the geometry of the VIPERS W1+W4 fields. The dark
matter haloes identified in the simulation have been populated with
galaxies using the halo occupation distribution (HOD) technique.
Because of the halo mass resolution of the simulation which is too large to
host the faintest galaxies observed in VIPERS, the method of
\citet{delatorre13b} has been applied to reconstruct haloes below the
resolution limit. Each halo has then been populated with galaxies
according to its mass as described by the HOD. The latter has been
calibrated directly on the VIPERS data as presented in
\citet{delatorre13}. To obtain fully realistic VIPERS mocks one needs
to reproduce the VIPERS survey selection function. This has been done
following several steps. First, the magnitude cut $i_{\rm AB}<22.5$
and the effect of the colour selection on the radial distribution of
the mocks have been applied. The mock catalogues thus obtained are
similar to the parent photometric sample used as target galaxy sample
for spectroscopy in VIPERS. The slit-positioning algorithm with the
same setting as for the data has further been applied to parent mock
catalogues. This allows us to reproduce the VIPERS footprint on the
sky, the small-scale angular pair incompleteness, and the variation of
TSR across the fields. Finally, random redshift errors has been added
to mock galaxy redshifts, similar to that present in the data.  This
procedure allows us to produce realistic mock galaxy catalogues that
contain the detailed survey completeness function and observational
biases of VIPERS.

\section{Galaxy clustering estimation} \label{sec:estimation}

We quantify galaxy clustering in redshift space by estimating the
anisotropic two-point correlation function $\xi(s,\mu)$, where $s$ is
the redshift-space separation of galaxy pairs and $\mu$ is the cosine
of the angle between the separation vector and the line of sight.  We
generate a catalogue of randomly distributed objects subject to the
same angular and radial selection as the true data, and use the
\citet{landy93} estimator:
\begin{equation}
\xi(s,\mu)=\frac{GG(s,\mu)-2GR(s,\mu)+RR(s,\mu)}{RR(s,\mu)}, 
\label{eq:xir}
\end{equation}
where $GG(s,\mu)$, $GR(s,\mu)$, and $RR(s,\mu)$ are respectively the
normalized galaxy-galaxy, galaxy-random, and random-random pair counts
in bins of $s$ ($\Delta(\log_{10}s)=0.1$) and $\mu$
($\Delta\mu=0.01$). This estimator has been shown to provide a nearly
unbiased estimate of the two-point correlation function, while
minimising its variance \citep{landy93}.  We typically use random
samples with 30 times more objects than in the true data, to reduce
their shot noise contribution to a negligible amount.

In this work we shall estimate the growth rate by fitting RSD models
not to the full shape of $\xi(s,\mu)$, but rather to its first two
even multipole moments, $\xi^{(0)}(s)$ and $\xi^{(2)}(s)$, defined as
\begin{equation}
\xi^{(\ell)}(s)=\frac{2\ell+1}{2}\int_{-1}^{+1}\xi(s,\mu)\mathcal{L}_\ell(\mu)d\mu,
\end{equation}
where $\mathcal{L}_\ell$ is the $\ell$-th order Legendre
polynomials. Such an approach is normally preferred in order to
prevent the size of data vectors and the resulting covariance matrix
from becoming too large for practical computation (but see
\citejap{mohammad16} for discussion of some drawbacks of this choice).

\section{Systematic selection effects}

The VIPERS angular selection function is the result of combining
several different angular completeness functions. Two of these are
binary masks (i.e. describing areas that are fully used or fully
lost). The first mask is related to defects in the parent photometric
sample (mostly areas masked by bright stars) and the other to the
specific footprint of VIMOS and how the different pointings are
tailored together to mosaic the VIPERS area.  These masks are easily
accounted for when defining the area and the auxiliary random samples
for clustering measurements.

A more complex selection is related to the incomplete target sampling
of VIPERS: on average 47\% of the targets satisfying the VIPERS
selection criteria can be placed behind a slit and observed, defining
what we call the average Target Sampling Rate (TSR).  In principle, we
should also account for the colour-colour pre-selection of the target
sample, which introduces a Colour Sampling Rate (CSR: see
\citejap{scodeggio16}). In practice, since the CSR can be safely
assumed to be constant over the survey area (thanks to the
particularly careful homogenization of the parent sample photometry --
see \citejap{guzzo14}), its effect is absorbed into the fit or model
describing the smoothed redshift distribution, as in
Eq.~(\ref{eq:nz}).  In any case, the CSR is consistent with being
unity for $z \ge 0.6$.  Finally, we have also to take into account how
the probability of measuring the redshift of a targeted galaxy depends
on observational conditions or technical issues (which can be
location-dependent), which we call the Spectroscopic Success Rate
(SSR). The relative relevance, modelling and overall impact of all
these effects is described in more detail the following sections.

\label{sec:angular_effects}

\subsection{Slit collisions}
\begin{figure}
\centering
\includegraphics[width=\hsize]{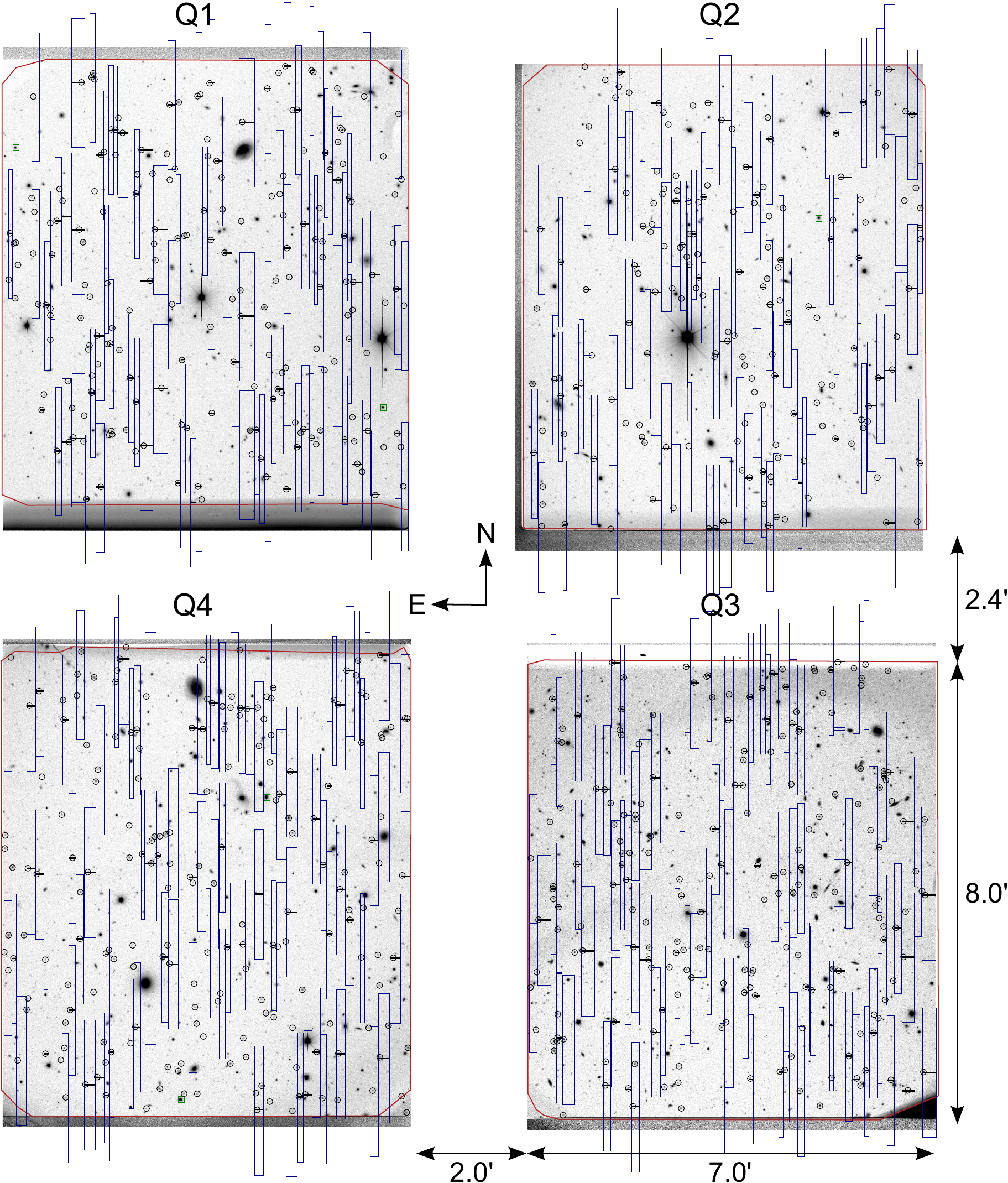}
\caption{Cartoon of the slit lay-out of a typical VIMOS pointing of
  the VIPERS survey (W1P082 in this case), superimposed on the actual
  DSS finding chart. The open circles with the tiny horizontal slits
  mark the target objects.  The vertical rectangles define the area
  where the resulting spectrum will fall, once the dispersing element
  (grism) is inserted.}
\label{fig:pointing}
\end{figure}

\begin{figure}
\centering
\includegraphics[width=.5\textwidth]{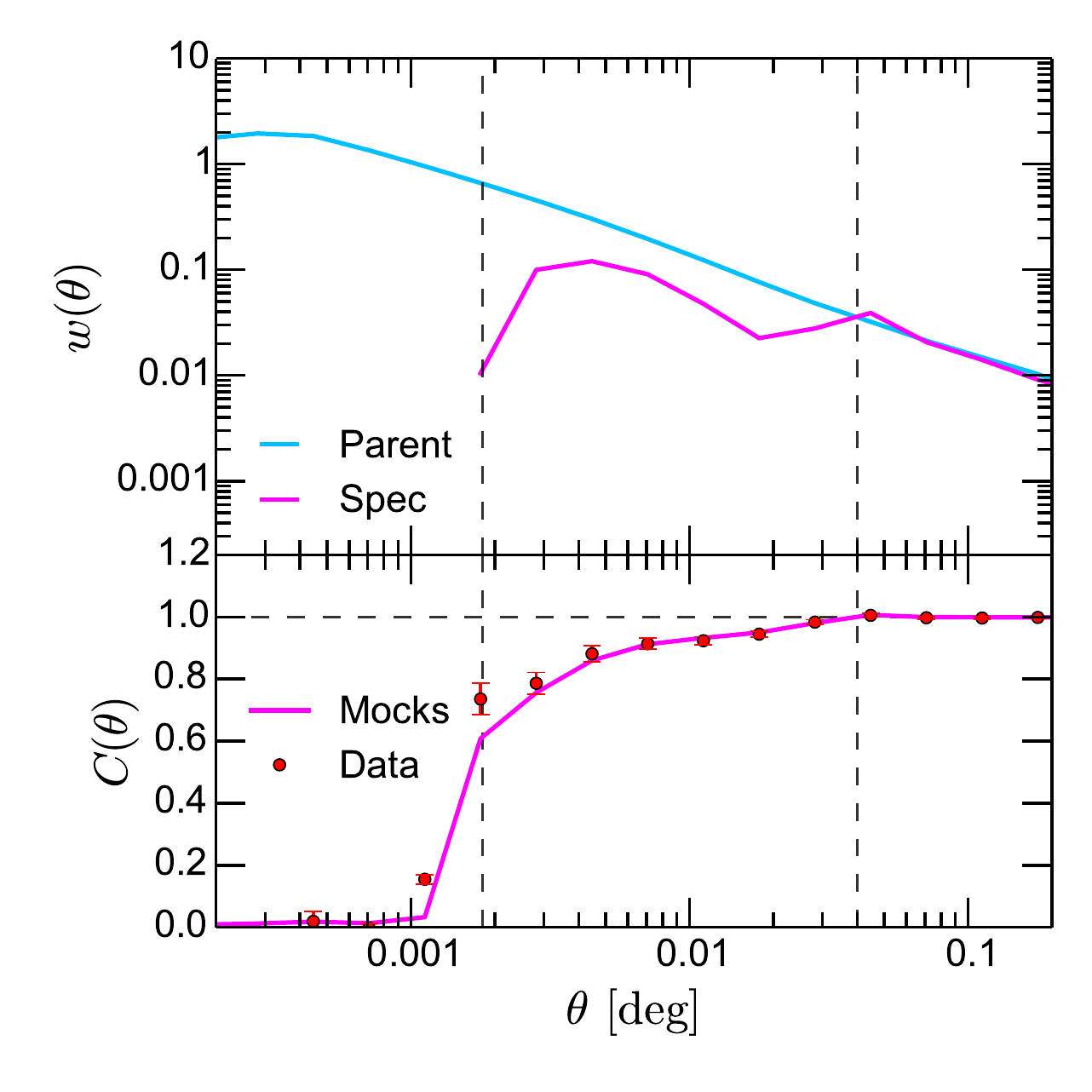}
\caption{Top: angular correlation function measured from the VIPERS W1
  mock samples. In order to enhance the signal-to-noise ratio, we
  display only the mean over 153 realisations. The angular correlation
  function of the parent/selected sample are displayed respectively
  with a cyan/magenta line. The two dashed vertical lines mark the
  typical angular size of the slits and the raw spectra. Bottom:
  completeness function, extracted from the mean of the 153 W1 mock
  samples (magenta line). The corresponding quantity measured from the
  VIPERS dataset is displayed with red circles.}
\label{fig:completeness_1d}
\end{figure}
\begin{figure}
\centering
\resizebox{0.93\hsize}{!}{\includegraphics{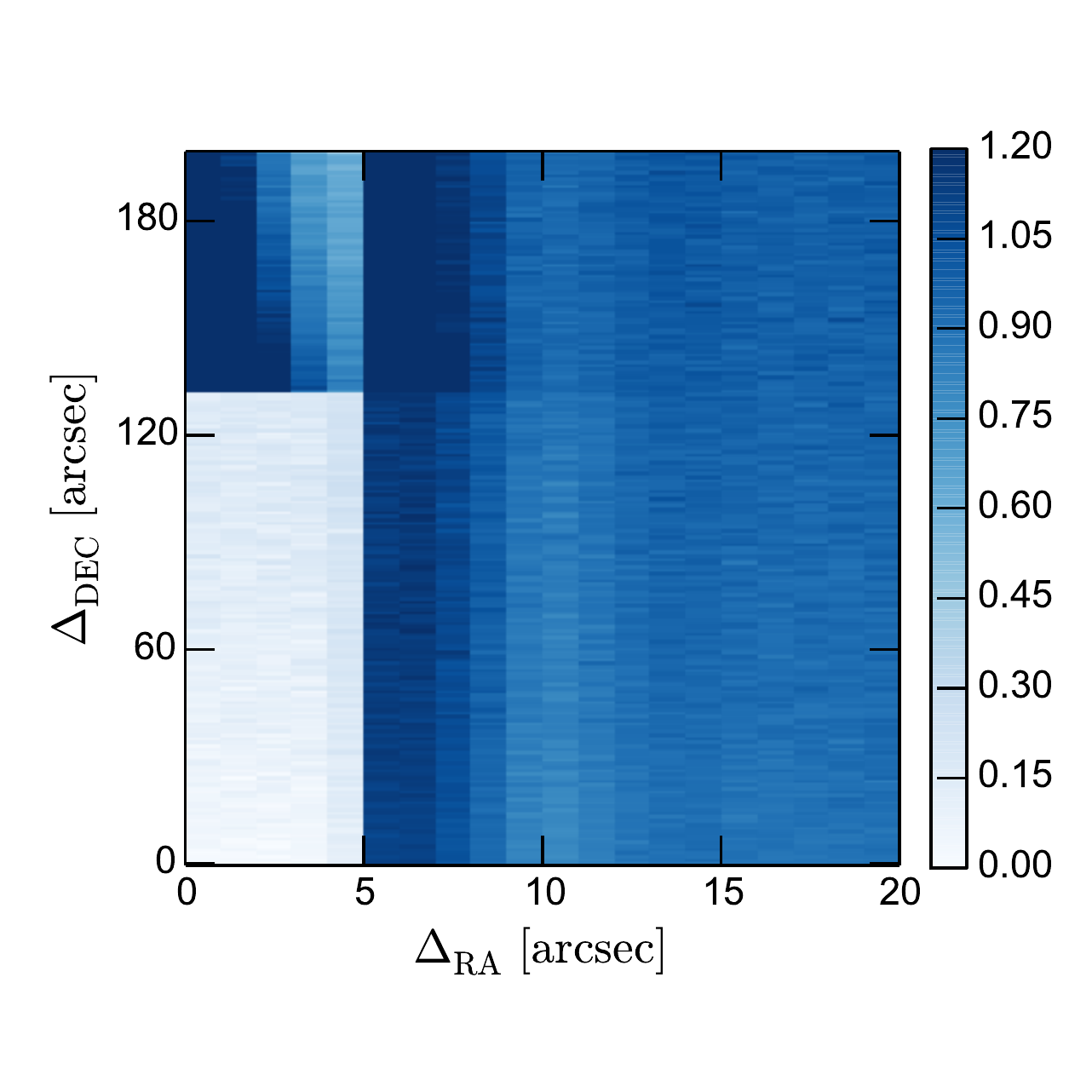}}
\caption{Two-dimensional completeness function, deprojected along the
  orthogonal coordinate axis RA-Dec. The imprint of a single galaxy
  spectrum is visible in the plot as the almost zero rectangular
  region at small angular separation. This region corresponds to the
  typical area occupied by the projected spectrum of an observed
  source. The enhancement of clustering in the top left region of the
  plot is produced by the particular displacement along common columns
  of slits within a quadrant. Note the very different scale of the
  abscissa and the ordinates.}
\label{fig:completeness_2d}
\end{figure}

A multi-object spectrograph survey must inevitably face the
limitations imposed by the mechanics of how light from the targets is
collected on the focal plane.  Either fibres or `slitlets' (as in the
case of VIMOS) impose a minimum physical size below which the spectrum
of two adjacent galaxies on the sky cannot be collected at the same
time. This suppresses completely the small-scale clustering amplitude,
unless multiple telescope visits of the same field are performed
(which is not the case with VIPERS).  Furthermore, the same limit on
close pairs causes high-density regions on the sky to be more poorly
sampled with respect to low-density regions; this introduces a
mismatch that, as we shall show, affects the amplitude of clustering
on all scales. In VIMOS, this effect is further enhanced by the
slit-positioning optimisation software \citep[SPOC:][]{bottini05},
which attempts to maximise the number of slits observed in each
quadrant and as such tends to homogenize the angular distribution of
targets.

Furthermore, in a multi-slit spectrograph such as VIMOS the dispersed
spectrum is imaged directly onto the detector.  As is evident from
Fig.~\ref{fig:pointing}, this creates another `forbidden zone'
perpendicular to the slit, where no other target can be observed
without causing two spectra to overlap (unlike in fibre spectrographs,
where fibres are typically taken away from the telescope to a standing
spectrograph and the spectra conveniently aligned and packed on the
CCD).  Since the projected length of the spectrum on the detector is
much larger than the corresponding size of the slit, this introduces
another typical scale below which the number of measured angular pairs
will be reduced, again limiting the sampling of overdensities on the
sky.  In VIPERS, the spectral dispersion is always oriented along the
North-South direction, so the depletion of galaxy pairs will be
anisotropic on the sky and will be larger along the declination
direction.

The impact of these effects on angular clustering is quantified in
Fig.~\ref{fig:completeness_1d}, where in the top panel we have plotted
for both the average of 153 mocks (solid lines) and the VIPERS data
(filled points) the angular correlation function of the parent and
spectroscopic samples ($w_p(\theta)$ and $w_s(\theta)$,
respectively). The bottom panel shows instead the ratio of the
corresponding numbers of pairs (bottom panel), defined as
\begin{equation}
C(\theta)=\frac{1+w_s(\theta)}{1+w_p(\theta)}\,\,\,\,  .
\label{eq:ctheta_formula}
\end{equation}
In this figure we find clear evidence of the two angular scales
discussed earlier, which are related to the width and length of the
spectra; these and identified in the figure by the vertical dashed
lines.  The origin of this effect can be better identified if we split
the separation angle $\theta$ into its components along the right
ascension and declination directions, $\Delta_{\rm RA}$ and
$\Delta_{\rm DEC}$.  The angular completeness map $C(\Delta_{\rm RA},
\Delta_{\rm DEC})$, corresponding to
Eq.~(\ref{eq:ctheta_formula}) is shown in
Fig. \ref{fig:completeness_2d}.  Here the `shadow' of the target
spectra is recognisable as the rectangular region with nearly zero
counts at small separations.  The few residual counts in this area are
produced by the small variations in the slit length, together with the
effect of the few serendipitous targets observed by chance within the
slit of a primary target.

Translated to spatial scales, this angular selection function results
in a strong suppression of the clustering amplitude below $1\mhmpc$,
as shown by the dotted line in Fig.~\ref{fig:syst_err_spoc_2} .  In
\citet{delatorre13}, we corrected for this effect by up-weighting each
galaxy-galaxy pair at a given angular separation $\theta_{ij}$ by the
inverse of the corresponding value of $C(\theta_{ij})$, i.e.
\begin{equation}
w^A(\theta)=\frac{1}{C(\theta_{ij})}.
\label{eq:wa_formula}
\end{equation}
We shall discuss the effectiveness of this weight together with the
correction of the large-scale effect of the TSR, at the end of the
next section.

\subsection{Larger-scale effects}

\begin{figure}[]
\centering
\resizebox{0.95\hsize}{!}{\includegraphics{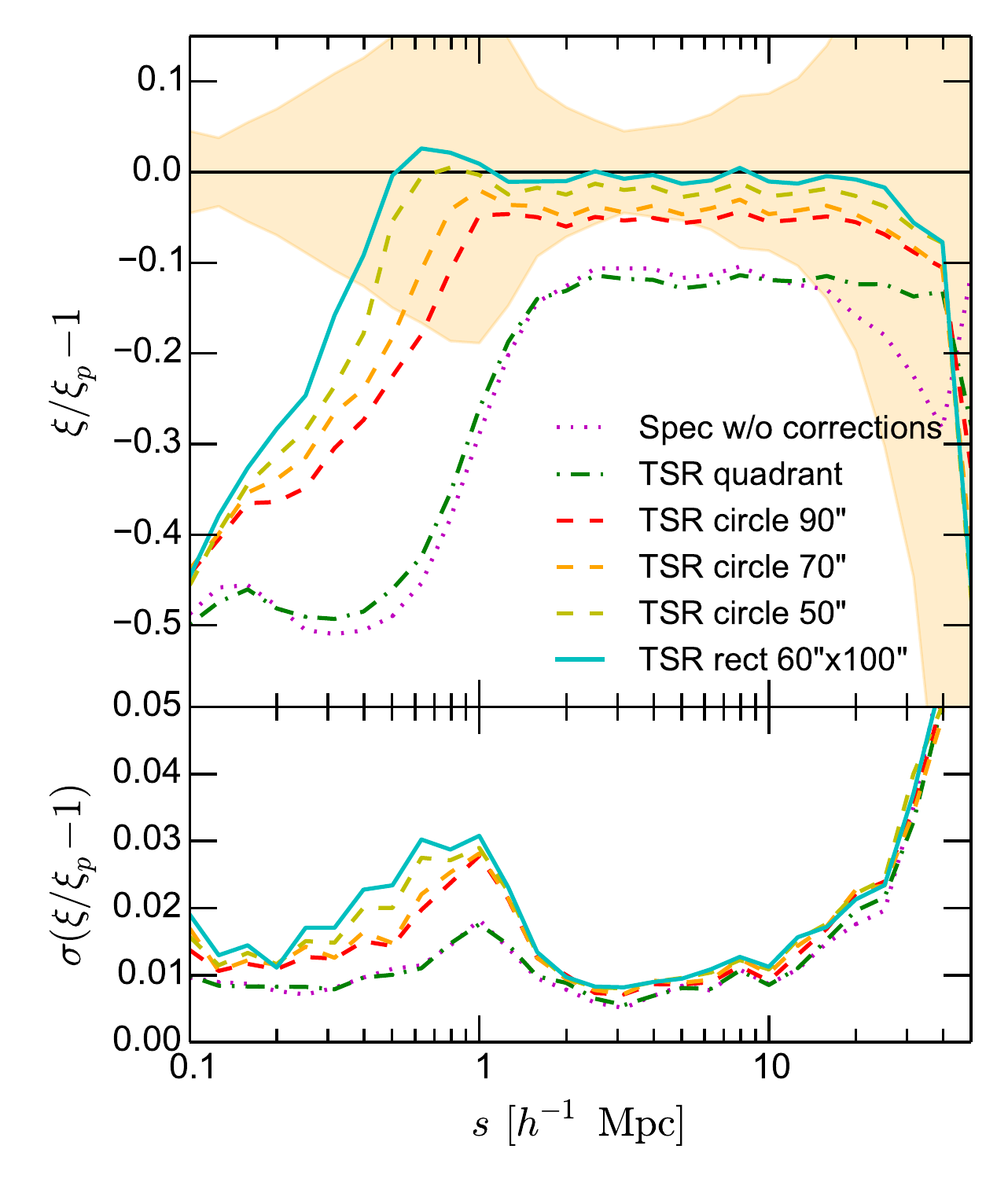}}
\caption{\footnotesize{Optimising the correction for the Target
    Sampling Rate on large-scales; the tests are based on the mean of
    153 mock samples. Top: systematic error on the real-space
    two-point correlation function introduced by the TSR (dotted
    line), confronted to the results of different strategies to
    estimate its local value and the corresponding weight (see text
    for details).  Circular apertures with varying radius ($r=90$,
    $70$ and $50\,\mathrm{arcsec}$), and a rectangular aperture
    $60\times100\,\mathrm{arcsec^2}$ are compared.  The dot-dashed
    line also shows the result of using a weight based only on the
    quadrant-averaged TSR.  Note that here the small-scale further
    correction based on Eq.~(\ref{eq:wa_formula}) has not been applied
    yet. Bottom: corresponding scatter of the different
    corrections. To allow comparison with the systematic error, this
    is also reported, for the rectangular aperture, as the shaded area
    in the top panel.}}
\label{fig:syst_err_spoc_2}
\end{figure}

Along with the drastic suppression at small separations, the physical
size of the slits is responsible for the inhomogeneous sampling
between high- and low-density regions across a single VIMOS
quadrant. This translates in an almost constant suppression of the
clustering amplitude on scales above $1\mhmpc$. The correcting scheme
we discuss here builds upon the original approach of
\citet{delatorre13}, in which galaxies are assigned a further weight
\begin{equation}
w_i=\frac{1}{\mathrm{TSR}_i}. 
\label{eq:tsr_weight}
\end{equation}
In that paper, however, the TSR used for each galaxy was simply the
average value over the corresponding VIMOS quadrant; in this way, all
target galaxies in a quadrant were up-weighted by the same factor. As
shown by the dot-dashed curve in Fig.~\ref{fig:syst_err_spoc_2}, when
considering the real-space correlation function $\xi(r)$ this
procedure has limited effect (note however than when combined with the
$w^A(\theta)=1/{C(\theta_{ij})}$ small-scale boost, it provides a
better correction: see Fig. 8 of \citejap{delatorre13}).

The improved correction adopted here uses instead a local estimate of
the TSR$ _i$, defined as the ratio of the local surface densities of
target and parent galaxies (i.e. before and after applying the target
selection); these are estimated as detailed below and then averaged
within an aperture of a given shape and size.  If we call these
quantities $\delta_i^p$ and $\delta_i^s$, the TSR$_i$ is defined as
\begin{equation}
\mathrm{TSR}_i=\frac{\delta_i^s}{\delta_i^p}.
\label{eq:tsr_formula}
\end{equation}
The continuous $\delta$ fields are obtained, starting from the
discrete distributions of parent and target galaxies, using a Delaunay
tessellation \citep{delaunay34} to estimate the density at the
position of each galaxy, and then linearly interpolating. These two
continuous fields are then used to compute the values of $\delta_i^p$
and $\delta_i^s$ within an aperture of a given shape and size.

We identified the best-performing geometry for this aperture through
the tests shown in Fig. \ref{fig:syst_err_spoc_2}.  The overall
correction is remarkable, since we are able to accurately recover the
parent $\xi(r)$ at large separations, both with a circular and a
rectangular aperture.  The rectangular aperture is the one providing
the best correction to real-space clustering, which can be understood
in terms of the anisotropy of the spectral `shadows' discussed
earlier.  The optimal size of the rectangular aperture is found to be
$60\times 100\,\mathrm{arcsec^2}$. The resulting
distribution of the TSR$_i$ values over the survey regions is shown in
Fig. \ref{fig:vipers_field}.

\begin{figure}
\hspace{-0.25cm}
\resizebox{\hsize}{!}{\includegraphics{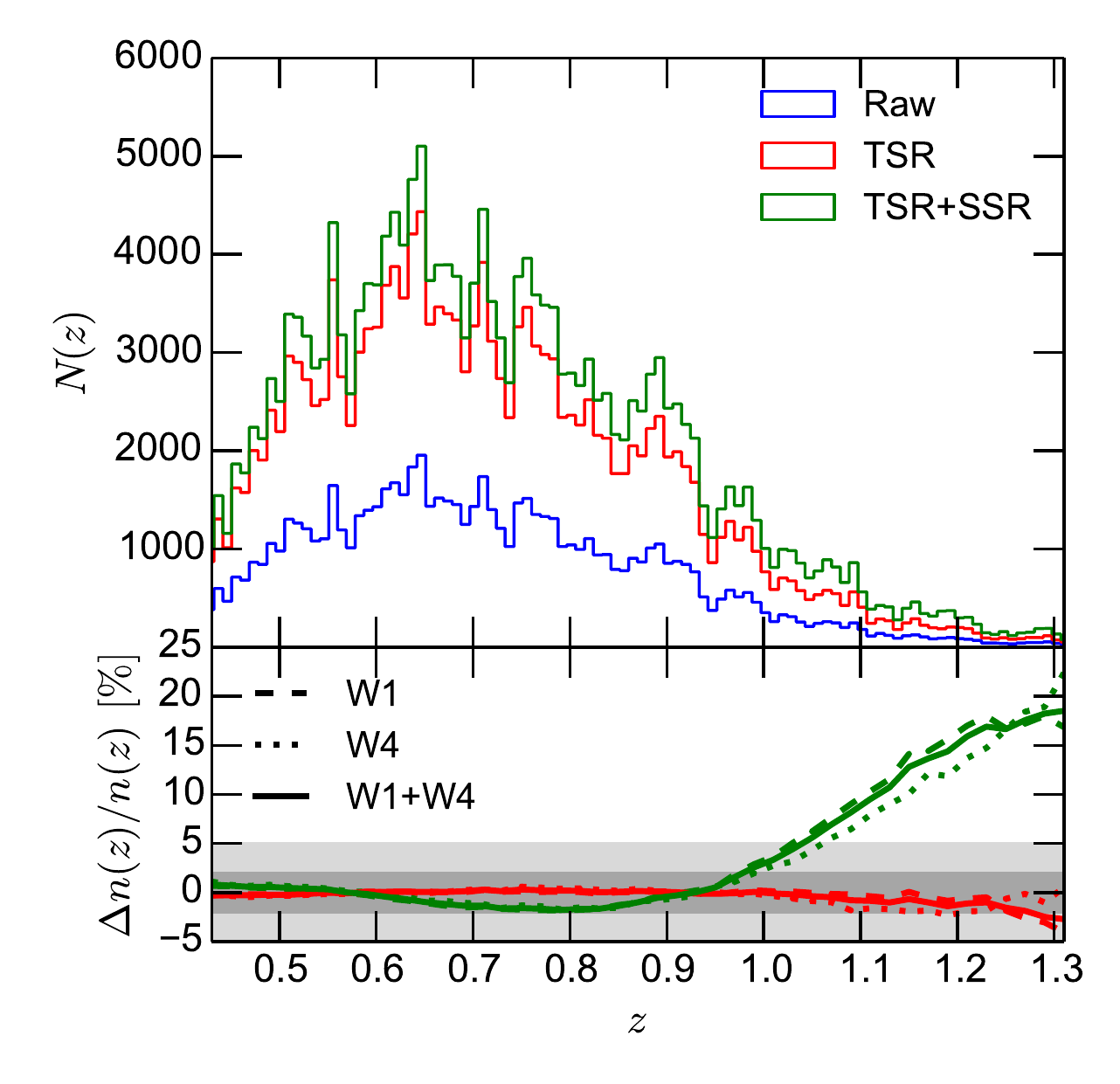}}
\caption{Impact of the Target Sampling Rate and the Spectroscopic
  Success Rate on the radial profile of the VIPERS galaxy samples. In
  the bottom panel we plot the relative difference of the $V_{\rm
    max}$ fits to the redshift distribution after applying the
  correction, to the same obtained from the observed
  histogram. Dashed, dotted and solid lines give the results for W1,
  W4 and the combined measurement, respectively. The smoothed radial
  profile is estimated using the $V_{\rm max}$ method. While the TSR
  does not affect the redshift distribution, the SSR enhances the
  number counts at $z>0.95$.}
\label{fig:nz_weights}
\end{figure}

\subsection{Redshift dependence of angular corrections}

Some of the corrections for angular selection biases do have an effect also on the redshift distribution.  Fig. \ref{fig:nz_weights} shows the effect of
correcting for the TSR and SSR on the observed redshift distribution
of the VIPERS data.  While the TSR does not introduce a significant
redshift dependence, the application of the SSR boosts the expected
number of galaxies in the distant ($z>1$) part of the sample.  This
clearly reflects the increased inefficiency to measure redshifts for
more and more distant objects.  To be fully consistent with the data,
then, the random samples used for the clustering analyses will have to
be weighted accordingly.

\section{Two-point correlations from the VIPERS data}

\begin{figure}
\centering
\includegraphics[width=.49\textwidth]{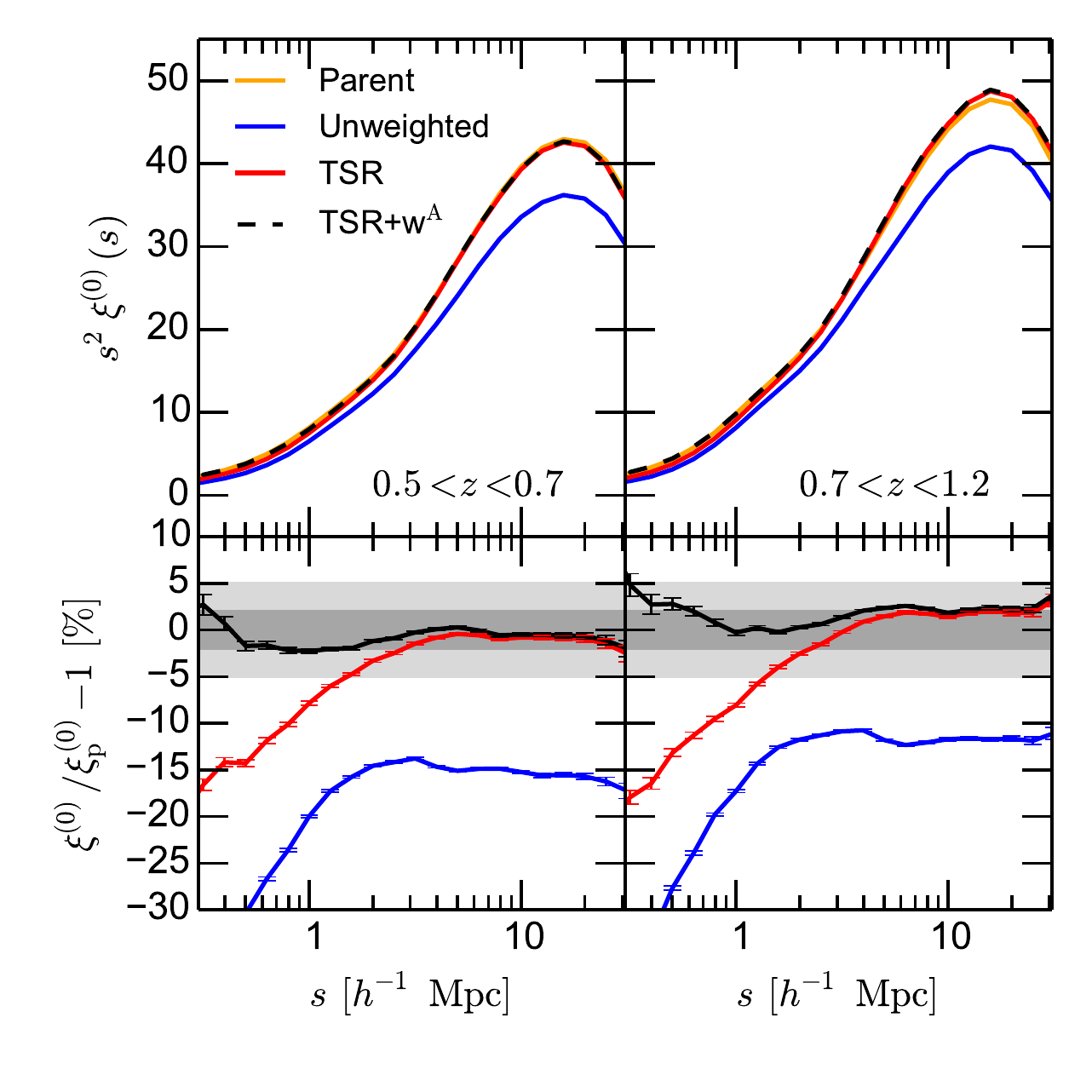}
\includegraphics[width=.49\textwidth]{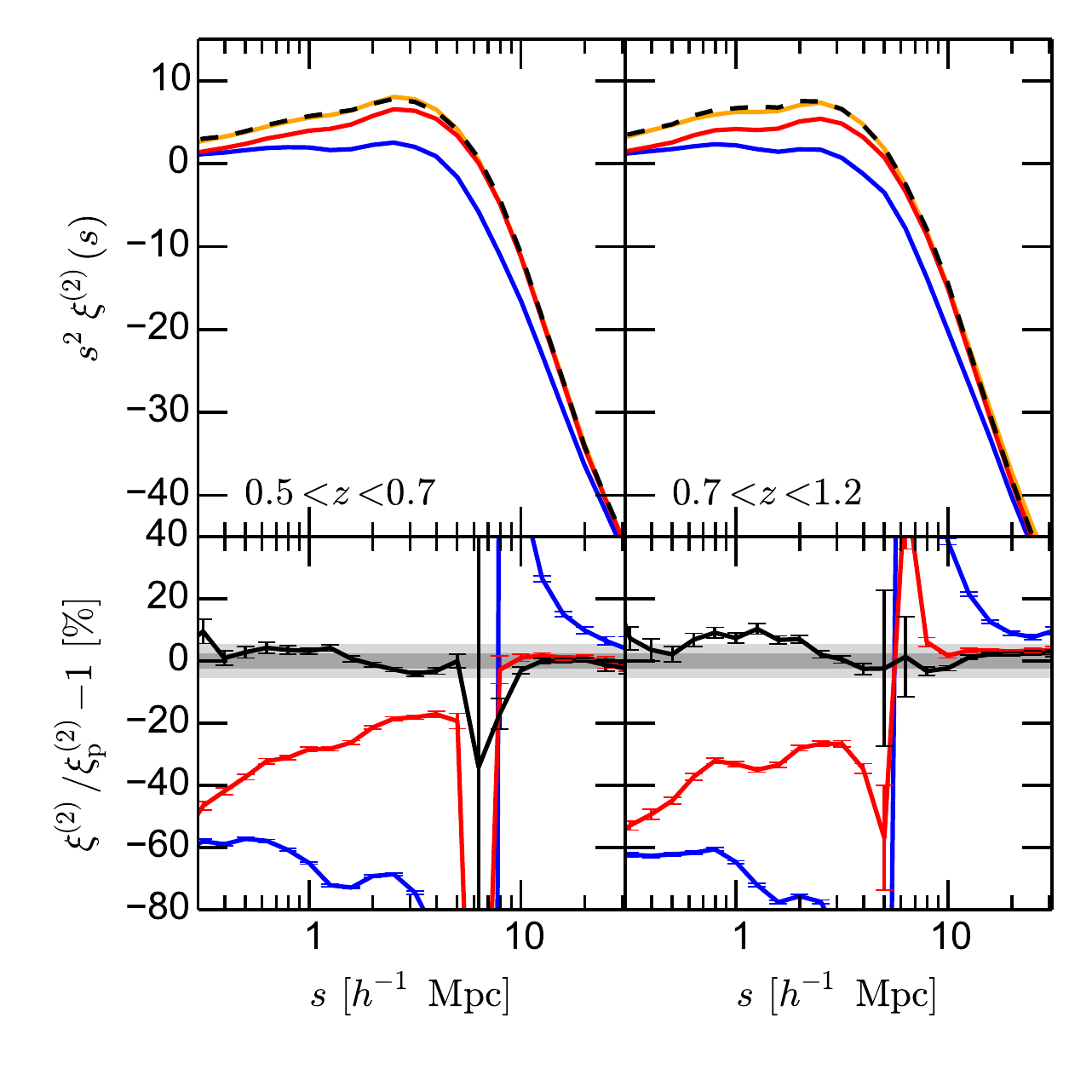}
\caption{Impact of the target selection effects and their correction
  on the amplitude of the monopole (left) and quadrupole (right) of
  the redshift-space correlation function $\xi(s,\mu)$.  Considering
  the mean over 153 mock samples, in the bottom panel we plot the
  fractional deviation of the multipoles measured using the observed
  sample from those obtained using the parent catalogue.}
\label{fig:syst_err_spoc_4}
\end{figure}

\begin{figure}
\centering
\resizebox{.95\hsize}{!}{\includegraphics{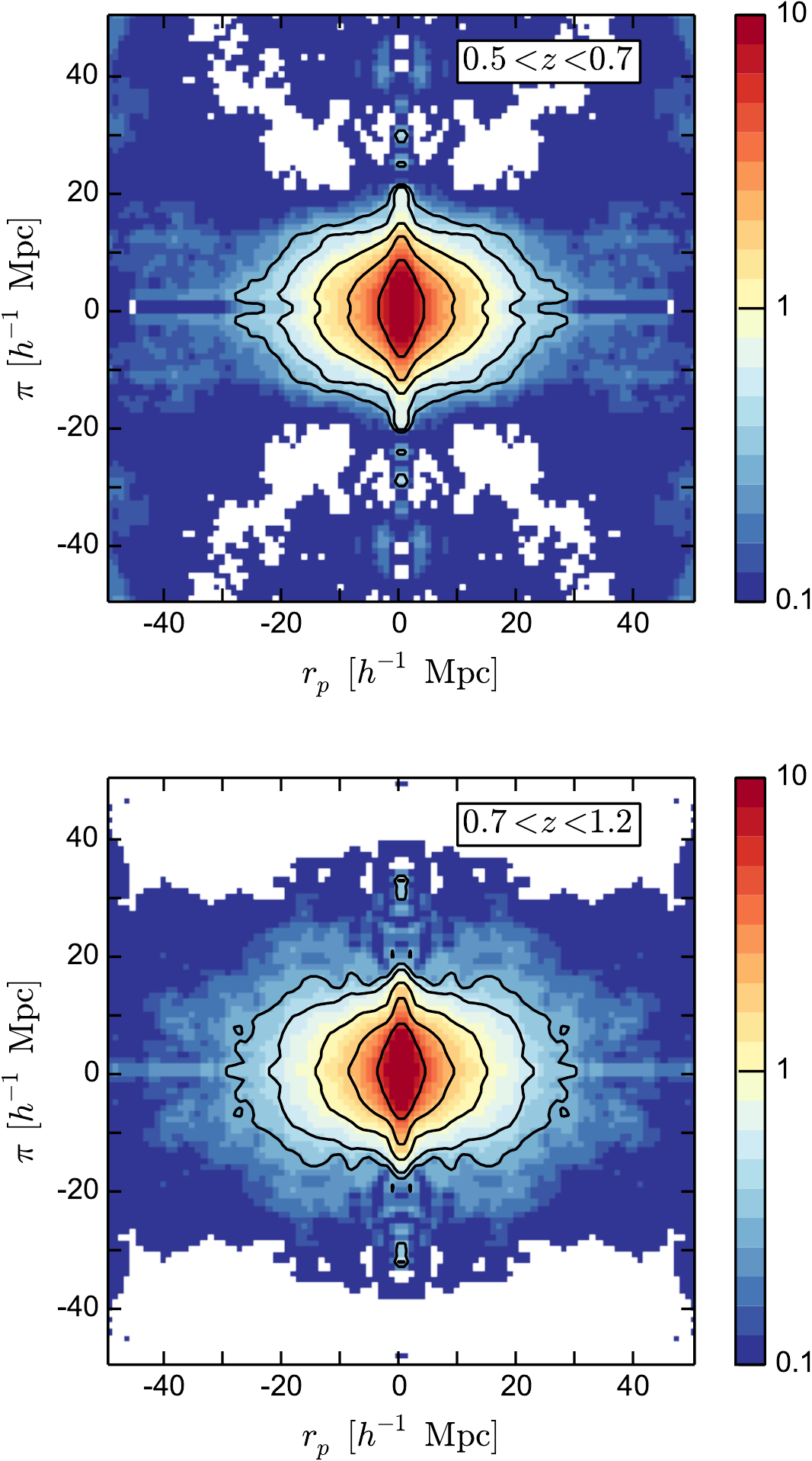}}
\caption{Final measurements of the anisotropic redshift-space
  correlation function, $\xi(r_p,\pi)$ from the final data of the
  VIPERS survey, within the two redshift ranges indicated by the
  labels.  Solid contours correspond to iso-correlation levels of 0.3,
  0.5, 1, 2, 5.  }
\label{fig:xip_data}
\end{figure}

We thus proceed to estimate the redshift space correlation function
and its moments for the VIPERS survey, adopting the weighting scheme
discussed in the previous sections, which we recap for convenience:
\begin{itemize}
\item each galaxy is upweighted by the inverse of its TSR defined by
  Eqs.~(\ref{eq:tsr_weight}) and (\ref{eq:tsr_formula}),
  $w^{\TSR}_i$, as well as by the inverse of its SSR, $w^{\SSR}_i$,
\item each galaxy-galaxy pair with angular separation $\theta$ is
  upweighted by the angular weight $w^A(\theta)$ defined in
  Eqs.~(\ref{eq:ctheta_formula}) and (\ref{eq:wa_formula}).
\end{itemize}

Pair counts in the two-point correlation function estimator of
Eq.~\ref{eq:xir} are then expressed as
\begin{flalign}
&GG(s,\mu)=\sum_{i=1}^{N_G}\sum_{j=i+1}^{N_G}w^A(\theta_{ij})w^{\TSR}_iw^{\TSR}_jw^{\SSR}_iw^{\SSR}_j\Theta_{ij}(s,\mu),\\
&GR(s,\mu)=\sum_{i=1}^{N_G}\sum_{j=1}^{N_R}w^{\TSR}_iw^{\SSR}_i\Theta_{ij}(s,\mu), \\
&RR(s,\mu)=\sum_{i=1}^{N_R}\sum_{j=i+1}^{N_R}\Theta_{ij}(s,\mu) \,\,\, ,
\end{flalign}
where $\Theta_{ij}(s,\mu)$ is equal to unity for $\log(s_{ij})$ in
$[\log(s)-\Delta \log(s)/2,\log(s)+\Delta \log(s)/2]$ and $\mu_{ij}$
in $[\mu-\Delta \mu/2,\mu+\Delta \mu/2]$, and null otherwise.

The final performance of this weighting scheme on the recovered
monopole and quadrupole of the redshift space correlation function are
shown in Fig. \ref{fig:syst_err_spoc_4}, for the two redshift ranges
considered in the analysis. The combined correction recovers the
amplitude of the monopole at the $2\%$ level, down to the
$\mathrm{Mpc}$ scale, yielding a quasi-unbiased estimate of
$\xi^{(0)}(s)$ on all comoving scales that will be used for the RSD
fitting. As for the quadrupole, we are able to have a reliable
measurement of $\xi^{(2)}(s)$ ($<5\%$ deviation from the fiducial
value) down to a few $\mathrm{Mpc}$. This is an encouraging result:
any uncorrected anisotropy from selection effects would be in danger
of inducing a spurious contribution to the quadrupole, since this is
our main measure of anisotropy.

Fig.~\ref{fig:xip_data} shows the measurement of the anisotropic
correlation function $\xi(r_p,\pi)$ obtained from the full VIPERS data
at $0.5<z<0.7$ and $0.7<z<1.2$.  A bin size $\Delta s=0.5$ \hmpc has
been used in both $r_p$ and $\pi$ directions. We combine the results
coming from the two VIPERS fields W1 and W4 simply by summing up the
pair counts in each bin of separation and normalising for the total
number of objects.

\section{Covariance matrix and error estimation}

Given the intrinsic correlation among different bins of the two-point
correlation function (and consequently of its multipoles), it is
essential to obtain a reliable estimate of the covariance matrix to be
used during the fitting procedure.  The fit is carried out performing
a maximum likelihood analysis of the data given the RSD model, that
can be more easily described as the search throughout the parameter
space of the position minimising the likelihood function $\mathcal{L}$
defined as
\begin{equation}
-2\ln\mathcal{L}=\sum_{i=0}^{N_b-1}\sum_{j=0}^{N_b-1}(y^d_i-y^m_i){\Psi_{ij}}(y^d_j-y^m_j).
\end{equation}
Here the observable $y=(\xi^0,\xi^2)$ is the monopole-quadrupole
combined vector; $\Psi\equiv {C}^{-1}$ is the precision matrix (the
inverse of the covariance matrix); $N_b$ is the total number of data
points; and indices $d$ and $m$ stand respectively for data and model.

\begin{figure}
\centering
\resizebox{.9\hsize}{!}{\includegraphics{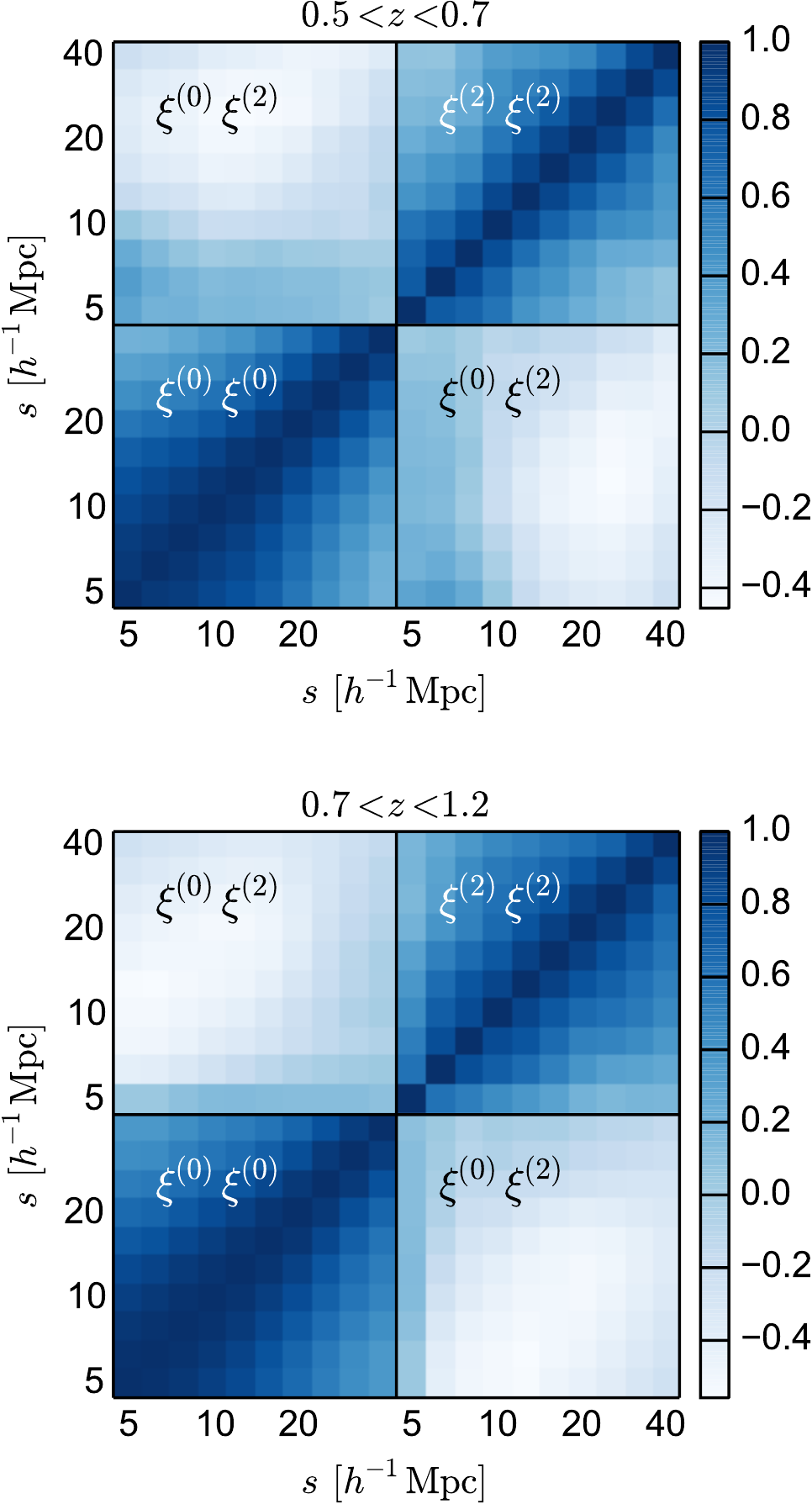}}
\caption{Correlation matrices for the combined monopole-quadrupole
  data vector, in the low (top) and high (bottom) redshift
  bin. Correlation matrices are computed as
  $R_{ij}=C_{ij}/\sqrt{C_{ii}C_{jj}}$, where $C$ is the covariance
  matrix estimated from a set of 153 independent mock samples. The
  bottom left and top right squares correspond respectively to the
  auto-covariance of the monopole $s^2\xi^{(0)}$ and the quadrupole
  $s^2\xi^{(2)}$, while the remaining squares show the
  cross-covariance terms. The scales under consideration range from
  $s_{\rm min}=5\mhmpc$ to $s_{\rm max}=50\mhmpc$ (from left to
  right).}
\label{fig:covariances}
\end{figure}

The covariance matrix ${C}$ is organised in four blocks corresponding
to the monopole-monopole, quadrupole-quadrupole and
monopole-quadrupole cross covariance (two identical blocks in the
latter case). The full monopole-quadrupole covariance matrix is
estimated from the 153 mock realisations as
\begin{equation}
\hat{C}_{ij}=\frac{1}{N_s-1}\sum_{k=1}^{N_s}\left(y^k_i-\bar{y}_i\right)\left(y^k_j-\bar{y}_j\right),
\end{equation}
where $N_s$ is the number of independent realisations used to estimate
the covariance, $y$ is the monopole-quadrupole vector, indices $i,j$
run over the data points and index $k$ runs over different
realisations. The mean value $\bar{y}$ is estimated by averaging the
measured values from different realisations, namely
\begin{equation}
\bar{y}=\frac{1}{N_s}\sum_{k=1}^{N_s}y^k.
\end{equation}
The corresponding correlation matrices obtained in this way for the two redshift sub-samples are shown in Fig.~\ref{fig:covariances}.

Given the large number of mock samples, the estimate and the inversion
of the covariance matrices can be achieved with good accuracy. However, the use of a finite number of mocks has two implications.
Firstly, the estimated precision matrix obtained by taking the inverse
of $\hat{C}$ is biased with respect to the true one, $\Psi$, with the
difference being well-represented by an inverse Wishart distribution.
Furthermore, the precision matrix $\Psi$ contains statistical errors
that propagate to the parameter space, affecting the derived errors on
the cosmological parameters. We follow \citet{percival14} and correct
for these effects by applying two correction factors. In the first
case, we can remove the systematic bias of the precision matrix by
rescaling $\hat C^{-1}$ as
\begin{equation}
\Psi=\left(1-\frac{N_{b}+1}{N_{s}-1}\right)\hat C^{-1}.
\end{equation}
The latter correction factor involves the total number of data points
$N_b$ and realisations $N_s$. It takes into account the typical
skewness characterising an inverse Wishart distribution and is capable
of providing an unbiased estimate of the precision matrix
\citep{hartlap07}. In the second case, the propagation of errors from
the precision matrix to the derived parameters can be corrected by
defining
\begin{equation*}
A=\frac{2}{(N_s-N_b-1)(N_s-N_b-4)},
\end{equation*}
\begin{equation}
B=\frac{(N_s-N_b-2)}{(N_s-N_b-1)(N_s-N_b-4)},
\end{equation}
and applying the correction factor
\begin{equation}
m_1=\frac{1+B(N_b-N_p)}{1+A+B(N_p+1)}
\end{equation}
to the estimated parameter covariance. In the previous equation, $N_p$
is the total number of free parameters.

\section{Modelling redshift-space distortions}

Redshift-space distortions arise because the apparent position of
galaxies is modified by the Doppler effect of their peculiar velocity
$\bs{v}$. In this way, the redshift-space position $\bs{s}$ of
galaxies located at $\bs{r}$ becomes
\begin{equation}
\boldsymbol{s}=\bs{r}+\frac{v_\parallel}{aH(a)}\bs{\hat{e}}_\parallel,
\end{equation}
where $a$ is the scale factor, $H(a)$ is the expansion rate and
$v_\parallel=\bs{v}\cdot\bs{\hat{e}}_\parallel$ is the component of
the galaxy peculiar velocity along the line of sight. Invoking mass
conservation, the redshift-space density field $\delta^s(\bs{s})$ can
be expressed as a function of its real-space counterpart
$\delta(\bs{r})$ as
\begin{equation}
\delta^s(\bs{s})=[1+\delta(\bs{r})] \bigg|\frac{d^3\bs{s}}{d^3\bs{r}}\bigg|^{-1}-1.
\label{eq:rsd_delta_1}
\end{equation}
The targeting of high-redshift galaxies in VIPERS means that the
largest pair separations are much smaller than the distance from the
observer, so we can use the small-angle plane-parallel approximation;
the Jacobian of the real-to-redshift space transformation then reduces
to
\begin{equation}
\bigg|\frac{d^3\bs{s}}{d^3\bs{r}}\bigg|=1-f\partial_\parallel u_\parallel,
\end{equation} 
where the normalized velocity field is defined as
$\bs{u}(\bs{r})=-\bs{v}(\bs{r})/[faH(a)]$. Substituting this
expression inside Equation (\ref{eq:rsd_delta_1}) it follows that
\begin{equation}
\delta^s(\bs{s})=\frac{\delta(\bs{r})+f\partial_\parallel u_\parallel}{1-f\delta_\parallel u_\parallel}.
\label{eq:rsd_delta_2}
\end{equation}
Taking the Fourier transform of this equation and making explicit the
dependence on $\mu={\bf \hat{k}\cdot\hat{r}}$, we obtain
\begin{eqnarray}
  \delta^{s}(k,\mu) &=& \int\frac{d^3\bs{s}}{(2\pi)^3} e^{-i{\bs{k}}\cdot{\bs{s}}} \delta^s(\bs{s}) \nonumber \\
  &=& \int\frac{d^3\bs{r}}{(2\pi)^3} e^{-i{\bs{k}}\cdot{\bs{r}}} e^{ik\mu f
  u_\parallel}\big[\delta({\bs{r}})+f\partial_\parallel u_\parallel\big].
\end{eqnarray}
The redshift-space power spectrum can thus be written as \citep{scoccimarro99}
\begin{eqnarray}
P^s(k,\mu) &=& \int\frac{d^3{\bs{r}}}{(2\pi)^3} e^{-i{\bs{k}}\cdot{\bs{r}}} \bigg\langle  e^{-ik\mu f \Delta 
  u_\parallel}\times\nonumber \\
&& \times\Big[\delta({\bs{x}})+ f\partial_\parallel u_\parallel\Big]\, 
\Big[\delta({\bs{x'}})+ f\partial_\parallel u_\parallel\Big]\,\bigg\rangle,   
\label{eq:rsd_pk} 
\end{eqnarray}
with $\Delta u_\parallel=u_\parallel(\bs{x})-u_\parallel(\bs{x'})$ and
$\bs{r}=\bs{x}-\bs{x'}$. This last equation completely describes the
anisotropies produced by peculiar velocities on the clustering of
matter particles at each separation. Here, the only assumption is the
plane-parallel approximation limit.

It is possible to identify two main regimes within which distortions
manifest themselves. At large separations, matter has a coherent flow
towards overdense regions. In this regime, the velocity field is
mainly irrotational \citep{bernardeau02} and can thus be described by
its divergence $\theta(\bs{x})=\bs{\nabla}\cdot\bs{u}(\bs{x})$. These
motions produce a systematic distortion of the large-scale
distribution along the line of sight. This `Kaiser effect'
\citep{kaiser87} is basically produced by the terms inside the square
brackets in Eq.~(\ref{eq:rsd_pk}).
 
In contrast, within the typical scale of haloes, galaxy orbits cross
each other: there is a random dispersion in velocities at a given
point, which convolves the redshift-space structure in the radial
direction.  The clustering amplitude is thus suppressed on small
scales, and structures appear stretched along the line of sight in the
so called `Fingers of God' \citep{jackson72}. This effect is mainly
generated by the exponential pre-factor involving the moment
generating function of the velocity field.

Eq.~(\ref{eq:rsd_pk}) is hard to use in its given form, because
we lack an analytic formula for the ensemble average term inside the
integral, particularly in the strongly non-linear regime.  But a
number of simpler approximate forms have been suggested, which aim to
provide a satisfactory representation of the redshift-space power
spectrum measured from galaxy surveys:

\noindent \textit{-- Kaiser model} \citep{kaiser87}: within the linear
theory approximation, the exponential pre-factor can be suppressed
since its impact on the largest scales is negligible and $\theta
\propto \delta$. If the galaxy-matter bias relation is also assumed to
be linear ($\delta_g=b\delta$), it follows that
\begin{equation}
P^s(k,\mu)=\Bigg(1+\frac{f}{b}\mu^2\Bigg)^2b^2P_{\delta\delta}(k),
\label{eq:kaiser_model}
\end{equation} 
where $P_{\delta\delta}=P$ is the linear real-space matter power
spectrum and $b$ is the linear galaxy bias.

\noindent \textit{-- Dispersion model} \citep{peacock94}: although the
previous model can reproduce the apparent enhancement of clustering at
large separations, it fails in the description of the non-linear
regime. The latter can be treated in a phenomenological way, by
artificially suppressing the linear theory predictions to account for
the effect of the Fingers of God. Eq.~(\ref{eq:kaiser_model}) can
thus be written as
\begin{equation}
P^s(k,\mu)=D\big(k\mu\sigma_{12}\big)\,\Bigg(1+\frac{f}{b}\mu^2\Bigg)^2b^2P_{\delta\delta}(k),
\label{eq:dispersion_model}
\end{equation}
where $D\big(k\mu\sigma_{12}\big)$ is an analytical damping
factor. This term depends on a nuisance parameter $\sigma_{12}$, which
plays the role of a pairwise velocity dispersion. The basic assumption
of the dispersion model is that $\sigma_{12}$ is not scale-dependent, but
rather can be fitted as a free parameter. An useful extension of this
model is to replace the linear $P_{\delta\delta}$ by a non-linear
version (using an analytic approximation such as HALOFIT). This then
allows the dispersion model to give the correct prediction for
$\mu=0$: such modes run transverse to the line of sight and undergo no
RSD effect. Note that some of the alternatives discussed here fail to
match the real-space power exactly at $\mu=0$: this is because they
are attempting the harder task of {\it predicting\/} the
non-linearities, rather than taking them from a fit to $N$-body
simulation data.

\noindent \textit{-- Scoccimarro model} \citep{scoccimarro04}: as soon as
the mildly non-linear regime is entered, the density and velocity
divergence fields must be treated separately to account for the
non-linear mode coupling between them. The ansatz proposed by
Scoccimarro is that the exponential pre-factor inside
Eq.~(\ref{eq:rsd_pk}) can be decoupled from the Kaiser term, so
that its impact on the clustering is limited only to the smallest
scales. In this case, it can be replaced with a damping factor similar
to the one already used in the dispersion model, leading to
\begin{equation}
P^s(k,\mu)=D\big(k\mu\sigma_{12}\big)\,\Big(b^2P_{\delta\delta}(k)+2fb\mu^2P_{\delta\theta}(k)+f^2\mu^4P_{\theta\theta}(k)\Big),
\label{eq:scoccimarro_model}
\end{equation} 
where $P_{\delta\theta}$ and $P_{\theta\theta}$ are respectively the
density-velocity divergence cross-spectrum and the velocity divergence
auto-spectrum.  When applying this (and the following) model to real
data, these quantities cannot be obtained from the data under
analysis. As such, applications of this (and the following) model have
used empirical fitting functions calibrated using numerical
simulations \citep{jennings11}.  In a parallel paper \citep{bel17},
new, more general formulas are proposed:
\begin{equation}
P_{\delta\theta}(k)=\bigg(P_{\delta\delta}(k)P^{lin}(k)e^{-k/k^*}\bigg)^{\frac{1}{2}},
\label{eq:fit_pdt}
\end{equation}
\begin{equation}
P_{\theta\theta}(k)=P^{lin}(k)e^{-k/k^*},
\label{eq:fit_ptt_1p}
\end{equation}
where $P^{lin}(k)$ is the linear matter power spectrum and $k^*$ is a
parameter representing the typical damping scale of the velocity power
spectra. This can be well described as
\begin{equation}
\frac{1}{k^*}=p_1\sigma_8^{p_2},
\end{equation} 
where $p_1,\,p_2$ are the only free parameters of the fit.
These forms for $P_{\delta\theta}$ and $P_{\theta\theta}$ have
valuable, physically motivated properties: they naturally converge to
$P_{\delta\delta}(k)$ in the linear regime, including a dependence on
redshift through $\sigma_8(z)$.

Full details on the derivation and performances of these fitting
formulas are presented in \citet{bel17}.  Their use in the analysis
presented in the following sections is a significant improvement over
previous applications of the Scoccimarro and TNS \citep{taruya10} models, as it allows
us to extend our tests to smaller scales and apply the models to high
redshifts as sampled by VIPERS.

\noindent \textit{-- Taruya (or TNS) model} \citep{taruya10}: the
non-linear mode coupling between the density and velocity divergence
fields is responsible for a systematic bias between measurements of
the power spectrum and its prediction using the previous RSD
model. The origin of this deviation is the additional terms inside
Eq.~(\ref{eq:scoccimarro_model}), which are not accounted for
within the previous ansatz. The corrected model can be written as
\begin{equation}
\begin{split}
 P^s(k,\mu)=D\big(k\mu\sigma_{12}\big)\,\Big(b^2P_{\delta\delta}(k)&+2fb\mu^2P_{\delta\theta}(k)+f^2\mu^4P_{\theta\theta}(k)+\\
 &+C_A(k,\mu,f,b)+C_B(k\,u,f,b)\Big),
\end{split}
\label{eq:taruya_model}
\end{equation}
where $C_A$ and $C_B$ are terms derived using perturbation theory,
which aim to account for the density and velocity divergence couplings
with the exponential pre-factor in Eq.~(\ref{eq:rsd_pk}).
See \citet{delatorre12} for the details of its application to biased tracers.
\begin{figure}
\centering
\includegraphics[width=0.48\textwidth]{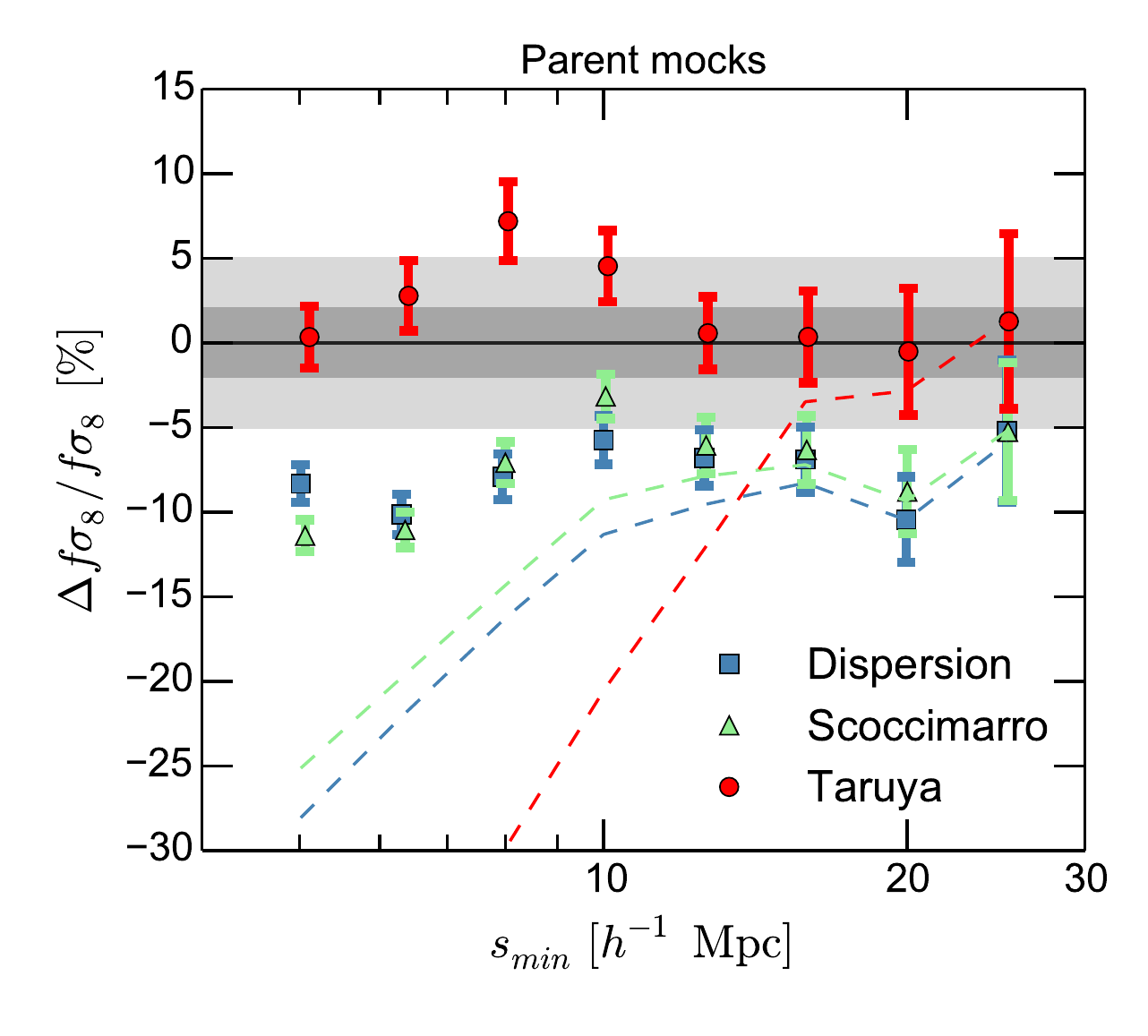}
\caption{Systematic errors on the measurement of the linear growth
  rate from the mean of 153 mock samples, using the three models
  discussed in the text. Here we used the parent mocks, to focus on
  the intrinsic performances of the models.  Relative systematic
  errors are plotted as a function of the minimum fitting scale
  $s_{\rm min}$.  $s_{\rm max}$ is always fixed at $50\mhmpc$.  The
  filled symbols correspond to the use of a Lorentzian form for the
  non-linear damping factor in the models, whereas dashed lines to a
  Gaussian one.}
\label{fig:syst_parent}
\end{figure}

All the tested RSD models feature a phenomenological damping factor
$D(k\mu\sigma_{12})$. The function $D(k\mu\sigma_{12})$ damps the
power spectra in the Kaiser term but also partially mimics the effects
of the pairwise velocity distribution in virialised systems. The
expected analytic form of the damping factor on large enough scales
assuming the Scoccimarro ansatz is Gaussian \citep{scoccimarro04}; but
analyses of simulated galaxy samples \citep{delatorre12} have shown
that a Lorentzian template provides a better practical fit.

Models in equations~\ref{eq:dispersion_model},
\ref{eq:scoccimarro_model} and \ref{eq:taruya_model} are all tested in
the next sections to understand their impact on the recovery of the
growth rate. In all cases, at each step of our Monte Carlo Markov chains we generate the full anisotropic
redshift-space power spectrum. For this we make use of
CAMB with the latest HALOFIT prescription for the non-linear
$P_{\delta\delta}$ \citep{takahashi12}, and Eqs.~\ref{eq:fit_pdt}
and \ref{eq:fit_ptt_1p} to generate the $P_{\delta\theta}$ and
$P_{\theta\theta}$ power spectra. The normalisation of the latter
real-space power spectra, which can be set by $\sigma_8$, is
degenerate with $f$ and $b$. This is why one generally parametrises
RSD models in terms $f\sigma_8$ and $b\sigma_8$ parameters. In the
case of the TNS model, however, this is not possible directly since
the $C_A$ term involves sub-terms that are not multiples of the
$f\sigma_8$ or $b\sigma_8$ parameters
\citep[e.g.][]{taruya10,delatorre12}. Therefore for the TNS model, and
for the others for consistency, we decide to treat $f$, $b$,
$\sigma_8$, $\sigma_{12}$ as free distinct parameters in the fit, and
provide derived constraints on $f\sigma_8$ a posteriori from the MCMC
chains.

It is important to emphasize that $\sigma_8(z)$ not only plays a role
in shaping the $C_A$ term, it also controls the level of non-linearity
in $P_{\delta\delta}$, $P_{\delta\theta}$, and $P_{\theta\theta}$. In
particular for $P_{\delta\delta}$, the HALOFIT non-linear correction
to the linear matter power spectrum is computed at each step of the
MCMC according to the tested value of $\sigma_8(z)$. This represents a
significant improvement over what is usually done in RSD analyses,
where $\sigma_8(z)$ is fixed to its fiducial value for the description
of $P_{\delta\delta}$.

In the end, we measure the Fourier-space multipole moments as
\begin{equation}
P^{(\ell)}(k)=\frac{2\ell+1}{2}\int_{-1}^{+1}P^s(k,\mu)\mathcal{L}_\ell(\mu)d\mu,
\end{equation}
and convert them to their configuration space counterparts as
\begin{equation}
\xi^{(\ell)}(s)=i^\ell\int\frac{dk}{2\pi^2}k^2P^{(\ell)}(k)j_\ell(ks),
\end{equation}
where $j_\ell$ denotes the spherical Bessel functions.

\section{Tests of RSD models}
\label{sec:model_tests}

\begin{figure}
\centering
\includegraphics[width=0.49\textwidth]{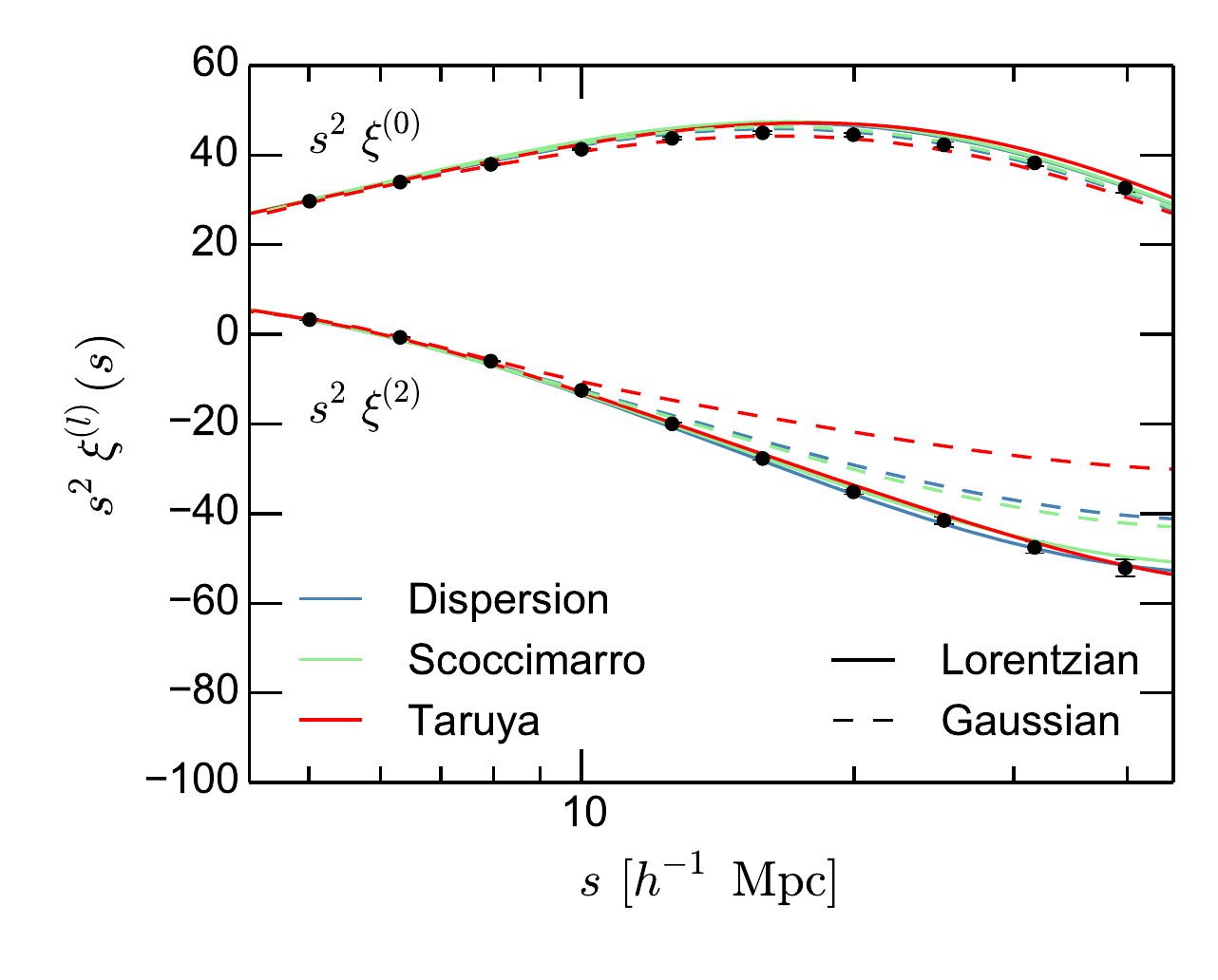}
\caption{Comparison between the best fit models for the monopole and
  quadrupole on the averaged parent mocks using different combinations
  of RSD models and damping factors. The fit uses down to $s_{\rm
    min}=5\mhmpc$. The use of a Gaussian damping in the models clearly
  dramatically worsen the accuracy of the fit, in particular for the
  large-scale quadrupole signal.}
\label{fig:comp_parent}
\end{figure}

We test in this section the RSD models introduced previously on our
set of $N_s=153$ mock catalogues. In practice, analysing each mock and
averaging the measurements would be computationally infeasible,
considering the large number of configurations to be tested.  We thus
chose to average the monopole and quadrupole measurements over the
mocks, scale the covariance matrix properly, and fit the models to
these average measurements. The aim is to reach a statistical
uncertainty that is a factor $\smash{1/\sqrt{N_s}}$ smaller than a
single VIPERS survey, to be able to detect potential systematics as
small as $1\%$. This process is more revealing and can show how well a
given model performs in recovering the detailed shapes of the
quadrupole and monopole correlation function.

We perform likelihood analyses of the mock mean measurements in
different configurations, starting the ideal case and moving on to
that in fully realistic conditions. All likelihood analyses are
carried out using an MCMC code, whose output has been cross-checked
with the independent MCMC code used in \citet{delatorre16}. We select
flat priors for the full set of free parameters, using boundaries that
allow a large set of late-time evolution cosmological models to be
considered as possible alternatives to standard $\Lambda$CDM. The full
list of priors is shown in Table~\ref{tab:priors}, while  the best-fit values for the parameters are listed in Table~\ref{tab:parameters}.  We vary the
minimum scale $s_{\rm min}$ of the fit to understand how to select the
best fitting range for the VIPERS data -- we expect all RSD models to
fail at sufficiently small and non-linear scales.  The maximum scale
of the fit is fixed at $s_{\rm max}=50\mhmpc$, above which errors on
the VIPERS measured monopole and quadrupole become too large.

\begin{figure}
\includegraphics[width=0.49\textwidth]{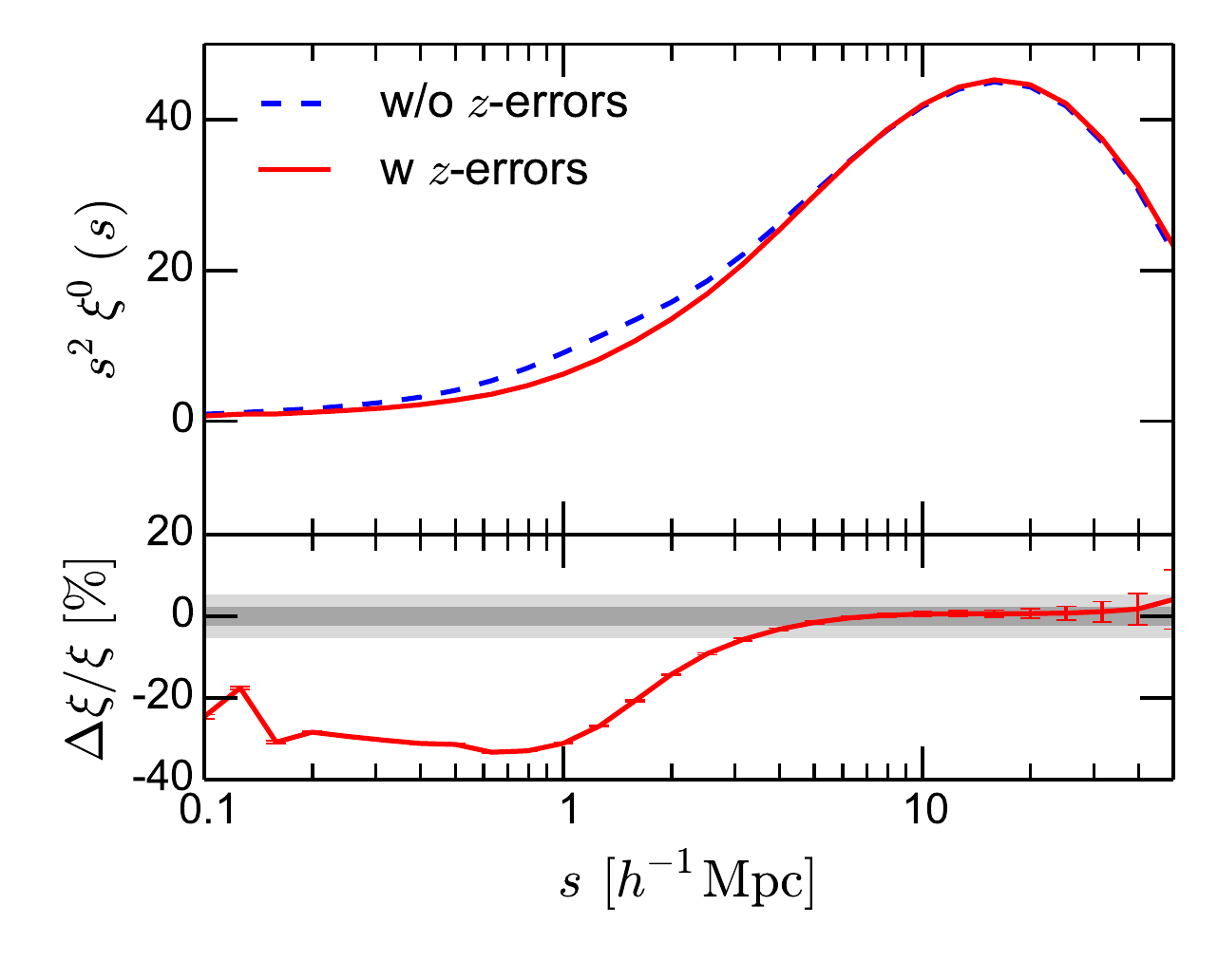}
\includegraphics[width=0.49\textwidth]{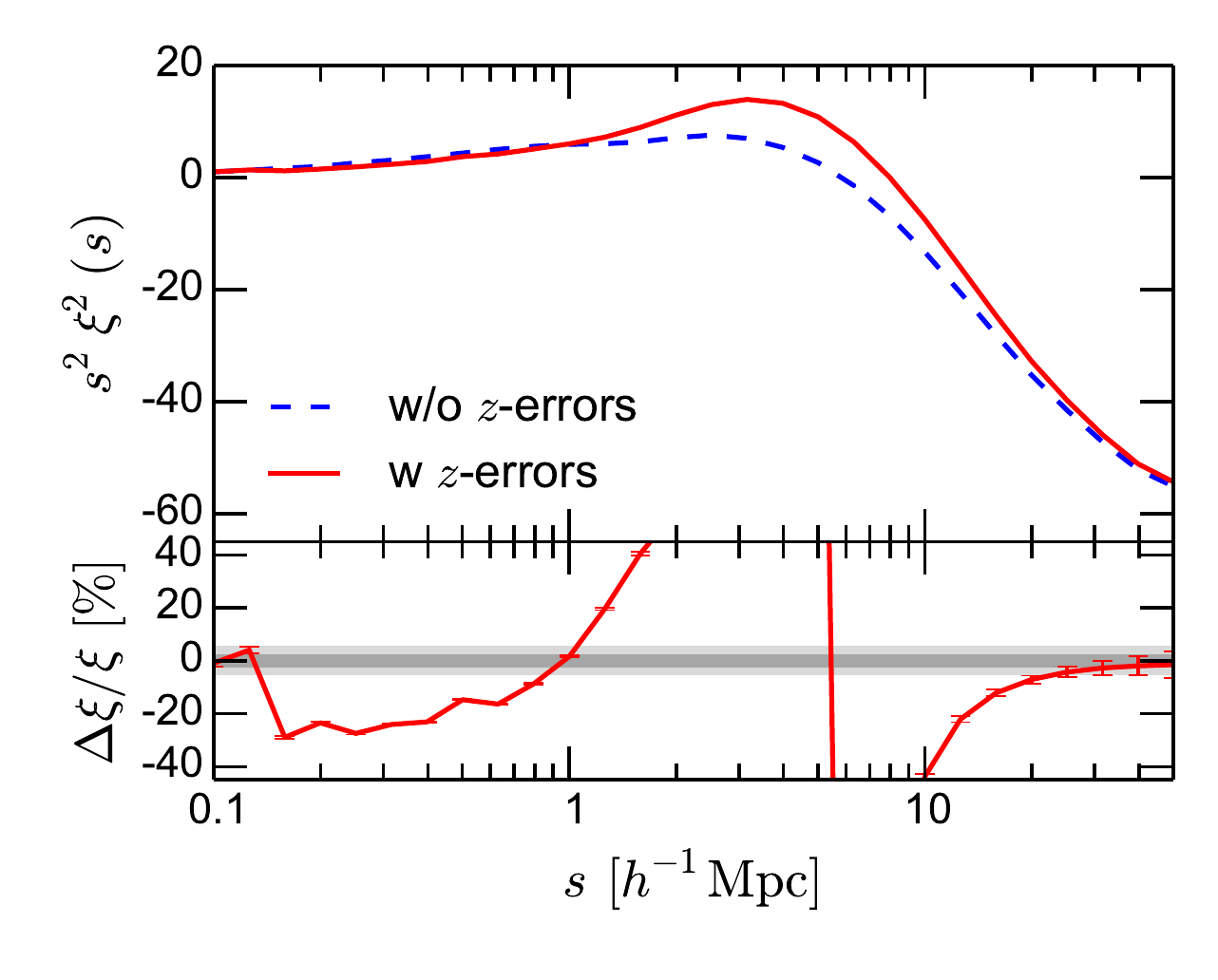}
\caption{Effect of redshift errors on the recovered monopole and
  quadrupole from the galaxy mocks, obtained by adding to the mock
  redshifts a random Gaussian deviate with dispersion equal to the
  {\sl rms } redshift error of the VIPERS.}

\label{fig:zerrors}
\end{figure}

\begin{figure}
\centering
\includegraphics[width=0.48\textwidth]{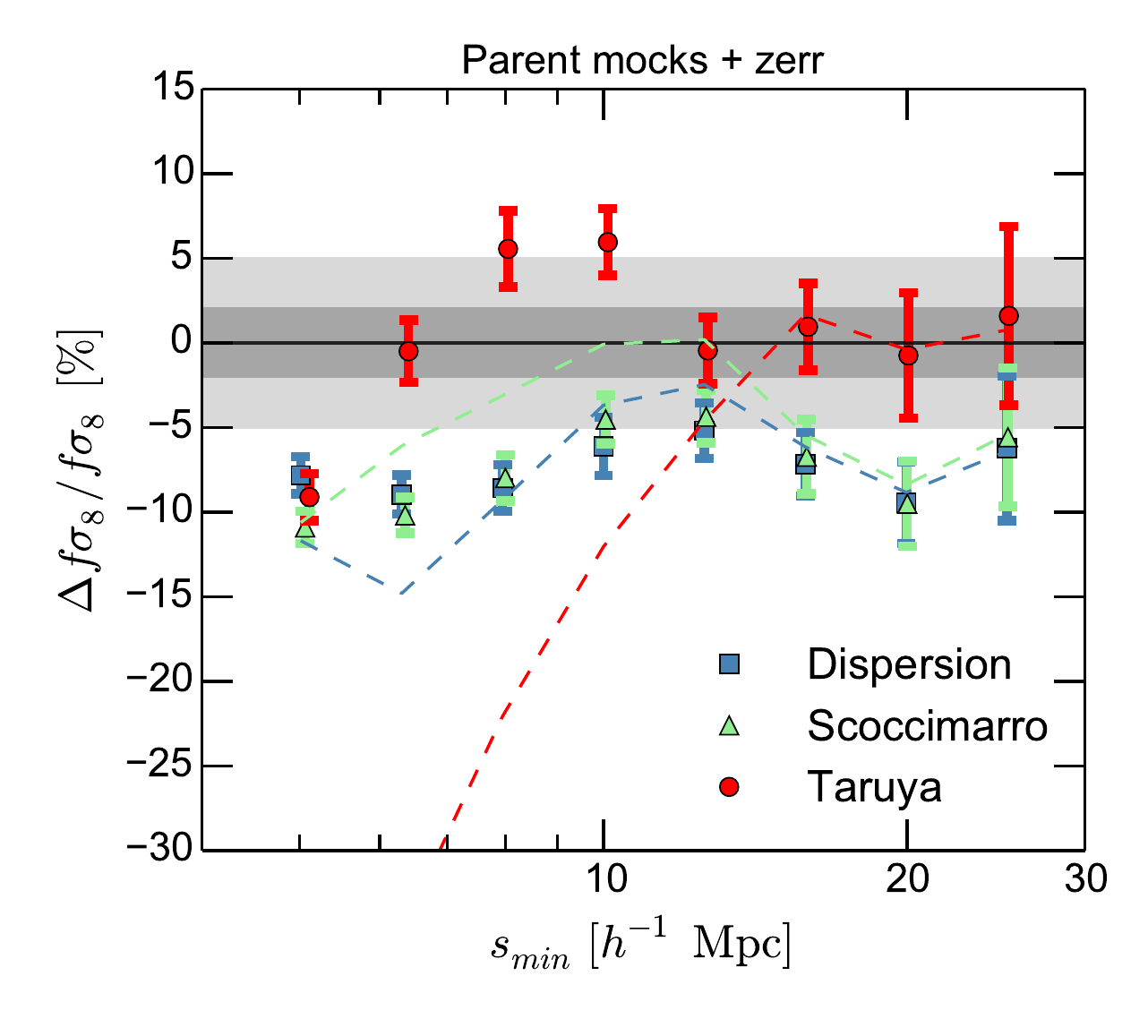}
\caption{Same as Fig.~\ref{fig:syst_parent}, but now including
  Gaussian redshift errors with dispersion equal to the {\sl rms }
  value measured for the VIPERS data, added to the mock galaxy
  redshifts.  Here the dashed lines correspond to the use of a
  Lorentzian damping only, which in Fig.~\ref{fig:syst_parent} was
  found to perform at best.  With redshift errors, this needs to be
  supplemented by a further Gaussian damping factor with dispersion
  fixed to the above {\sl rms} error value, to yield the values
  described by the filled symbols.}
\label{fig:syst_parent_errors}
\end{figure}

\subsection{Ideal case}

We first study the ideal case that neglects the complex VIPERS angular
selection by using the parent mocks. Here we consider mock
measurements from the full redshift range probed by VIPERS,
i.e. $0.5<z<1.2$, to avoid tuning the procedure towards small
systematic deviations on the two redshift bins. The relative
difference of the recovered $f\sigma_8$ with respect to the fiducial
one is shown in Fig. \ref{fig:syst_parent}. Redshift errors are not
considered here to understand how different RSD models behave in the
absence of any observational bias. Two types of small-scale damping
factor $D(k\mu\sigma_{12})$ are tested: the Lorentzian (filled
symbols) and Gaussian (dashed lines) forms. The overall trend of
models using Lorentzian damping favours the TNS model: it yields
almost unbiased measurements of the growth rate down to $s_{\rm min} =
5 \mhmpc$.  Some overestimation is however seen for minimum scales
close to $s_{\rm min}=8\mhmpc$, which in fact corresponds to the
zero-crossing scale of the quadrupole $\xi^{(2)}(s)$.

In contrast, both dispersion and Scoccimarro models consistently
underestimate the growth rate with an error that fluctuates between
5--10\%. Evidently, in all the cases the choice of a Lorentzian
damping yields smaller systematic deviations than with a Gaussian
damping.  This is reflected by the trend of the dashed lines, which
are close to the corresponding markers only when the minimum fitting
separation $s_{\rm min}$ is larger than $\sim 15\,\mhmpc$, while
rapidly deteriorating when smaller separations are included in the
fit. This is highlighted in Fig. \ref{fig:comp_parent}, where the
different best-fitting models for the monopole and quadrupole using
$s_{\rm min}=5\mhmpc$ are directly compared to the mock data. Using a
Gaussian damping, the model is no longer able to provide a good
description of $\xi^{(0)}$ and $\xi^{(2)}$. Actually, the fit is
mostly dominated by the small scales, whose data points have the
smallest errors, and this explains why separations above $7\mhmpc$ are
apparently the ones giving the strongest deviation between model and
data. This result is in close agreement with previous work on the
subject \citep[e.g.][]{bianchi12,delatorre12}.

\subsection{Case with redshift errors}

So far no redshift error was assumed in the mock samples. However,
real VIPERS redshifts have a significant uncertainty, which clearly
impact observed redshift-space distortions.  We know from the multiple
redshift measurements \citep{scodeggio16} that the redshift error
probability distribution for the VIPERS sample of reliable redshifts
used here, is well described by a Gaussian with standard deviation
$\sigma_z=0.00054(1+z)$.  This corresponds to a dispersion in galaxy
peculiar velocity of 163\,km\,s$^{-1}$.
 
\begin{figure}
\centering
\includegraphics[width=0.48\textwidth]{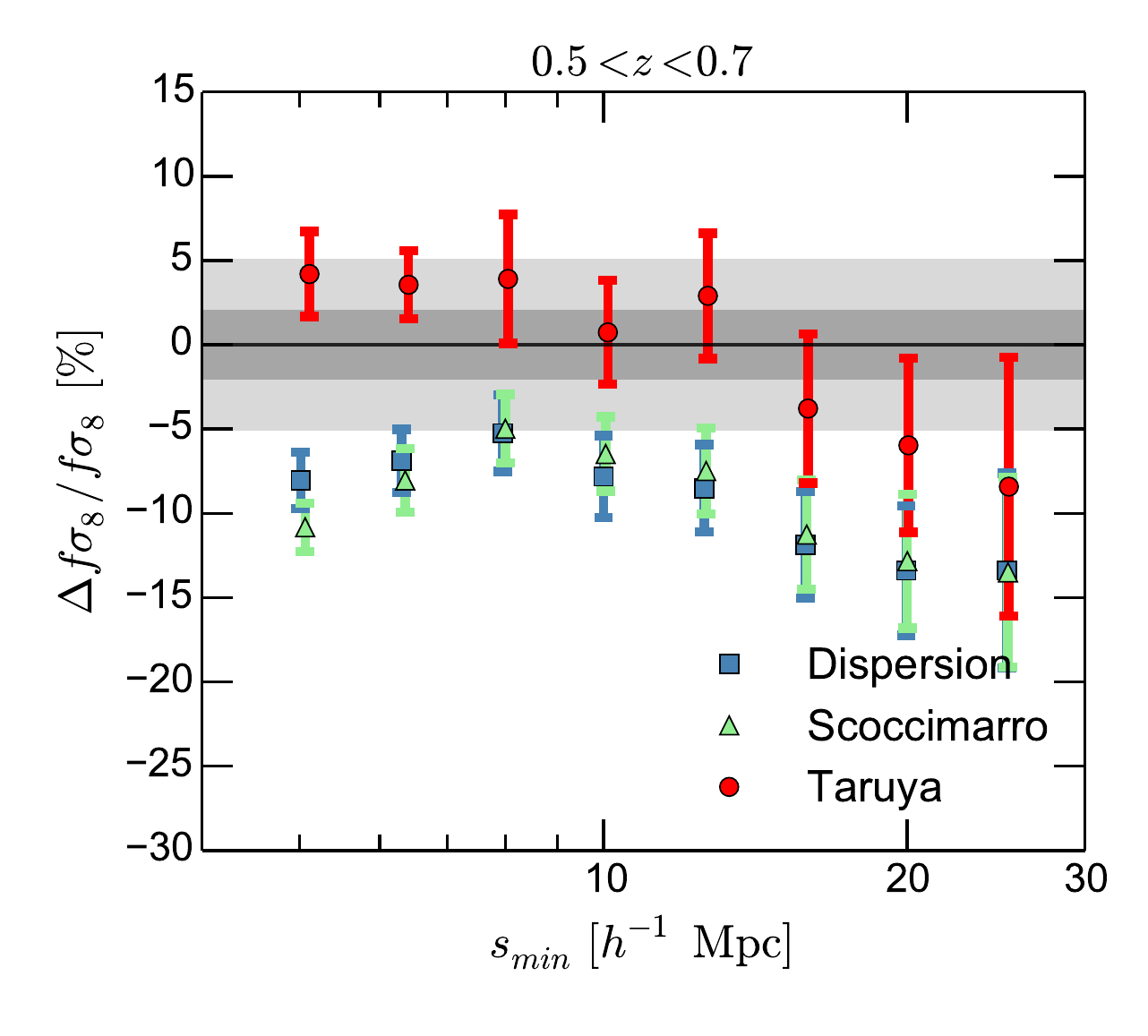}
\includegraphics[width=0.48\textwidth]{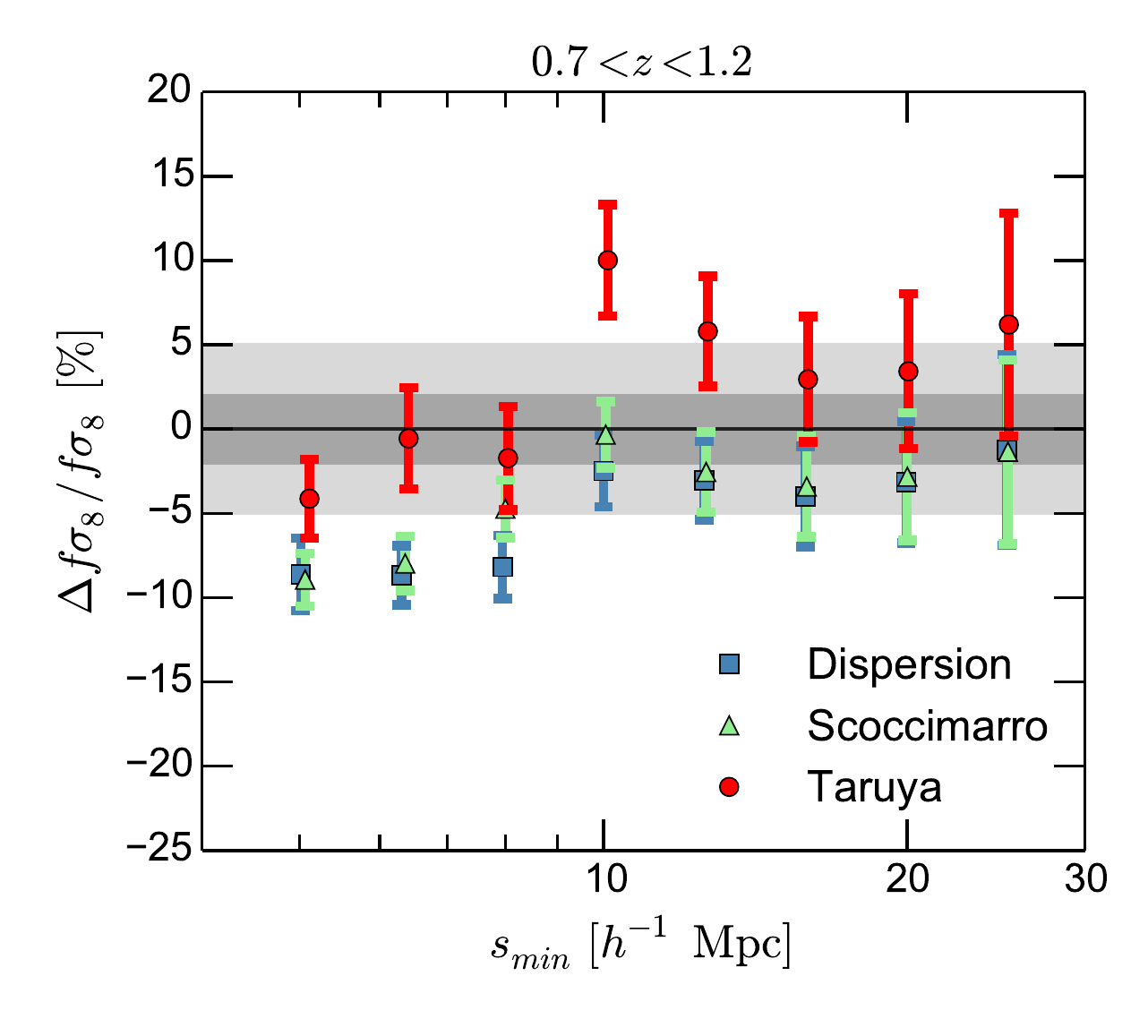}
\caption{The same as Fig.~\ref{fig:syst_parent_errors}, but now using
  the fully realistic `observed' mocks, on which all observational
  effects (masks, SPOC selection and redshift errors) have been
  included.  As before, error bars correspond to the error on the
  average of the 153 samples.  }
\label{fig:syst_obs_mocks}
\end{figure}

\begin{figure}
\resizebox{.9\hsize}{!}{\includegraphics{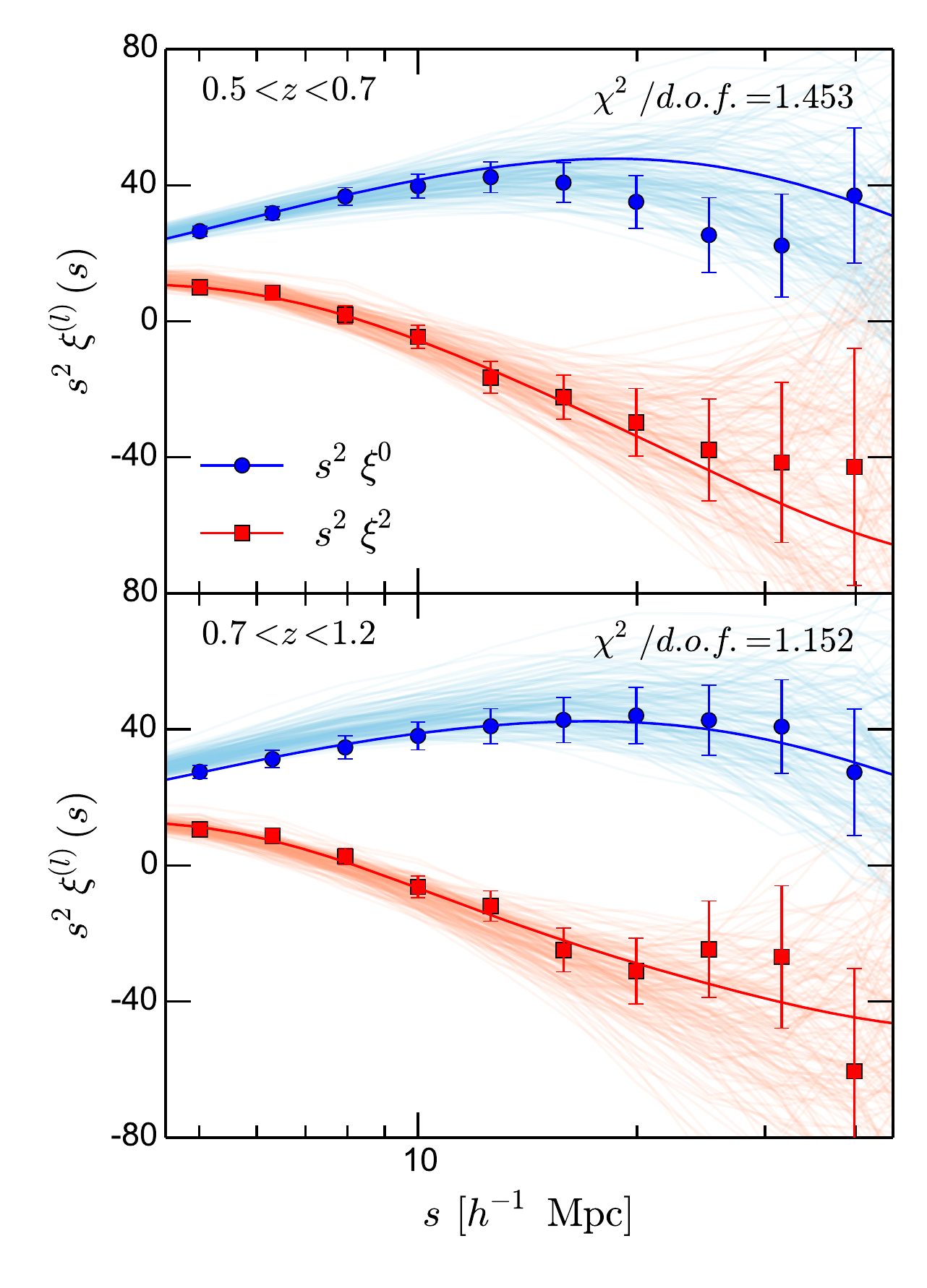}}
\caption{Monopole and quadrupole of $\xi(r_p,\pi)$ for the two
  redshift sub-sample of the final VIPERS dataset (solid points),
  together with the final best-fitting curves obtained using the TNS
  model, corresponding to the values reported in
  Table~\ref{tab:parameters}.  The likelihood computation has used
  data down to $s_{\rm min}=5\mhmpc$, as indicated by the tests.
  Error bars are $1-\sigma$ deviations, and correspond to the
  dispersion of the mock measurements.  Each of these is also shown as
  a faint background line. }
\label{fig:multipoles}
\end{figure}

By applying random errors to mock galaxy redshifts following the
VIPERS observed distribution, we can effectively see additional
distortions. These are shown in Fig. \ref{fig:zerrors}, where one can
see how the shapes of the monopole $\xi^{(0)}(s)$ and the quadrupole $\xi^{(2)}(s)$ are affected.
The imprint of redshift errors is similar to that of a small-scale
damping of the power spectrum. While the monopole
is damped below $4\mhmpc$, the quadrupole
 is corrupted over a range extending out to $\sim 20\mhmpc$. Clearly,
this effect needs to be carefully handled or modelled, if one hopes to
recover an unbiased value for the growth rate.  The consequences of
not correcting for this effect are shown by the dashed lines in
Fig.~\ref{fig:syst_parent_errors}, where we are repeating the tests of
Fig.~\ref{fig:syst_parent}, but now including redshift errors. As
feared, there is a significant deviation from the values of
$f\sigma_8$ previously measured with the models in the best
configuration, i.e. with the Lorentzian damping.

Rather than correcting for redshift errors in the measurements, as
performed for the angular selection selection, it is better to include
it in the modelling.  It is intuitive to supplement the models with a
convolution by an extra Gaussian distribution with standard deviation
fixed to the VIPERS {\sl rms } value of $\sigma_z=163~\rm{km}~s^{-1}$,
which corresponds to
\begin{equation}
\sigma_\pi=\frac{c\sigma_z}{H(z)},
\end{equation}   
in terms of line-of-sight comoving separation.
The filled symbols in Fig. \ref{fig:syst_parent_errors} show how with this extra damping
term we recover a performance similar to the more idealised case of Fig.~\ref{fig:syst_parent}.  

We therefore adopt the TNS model with Lorentzian damping and Gaussian
error damping, as our reference model for the final RSD analysis of
the VIPERS data. However, we will also test for consistency the
behaviour of the other two models on the actual data, to verify
whether the trends seen in the mocks are confirmed.

\subsection{Fully realistic case}

\begin{figure*}
\centering
\resizebox{.48\hsize}{!}{\includegraphics{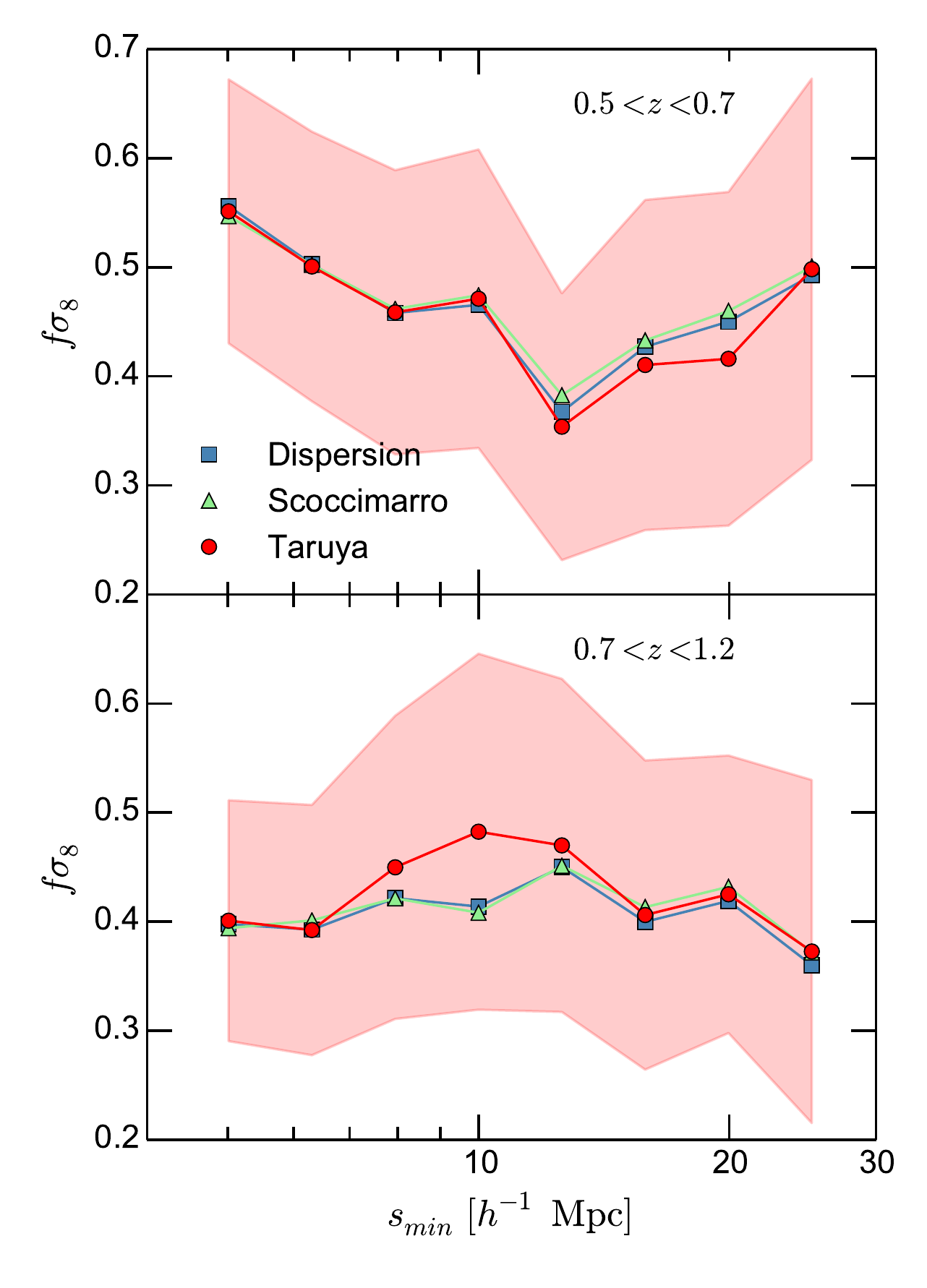}}
\resizebox{.48\hsize}{!}{\includegraphics{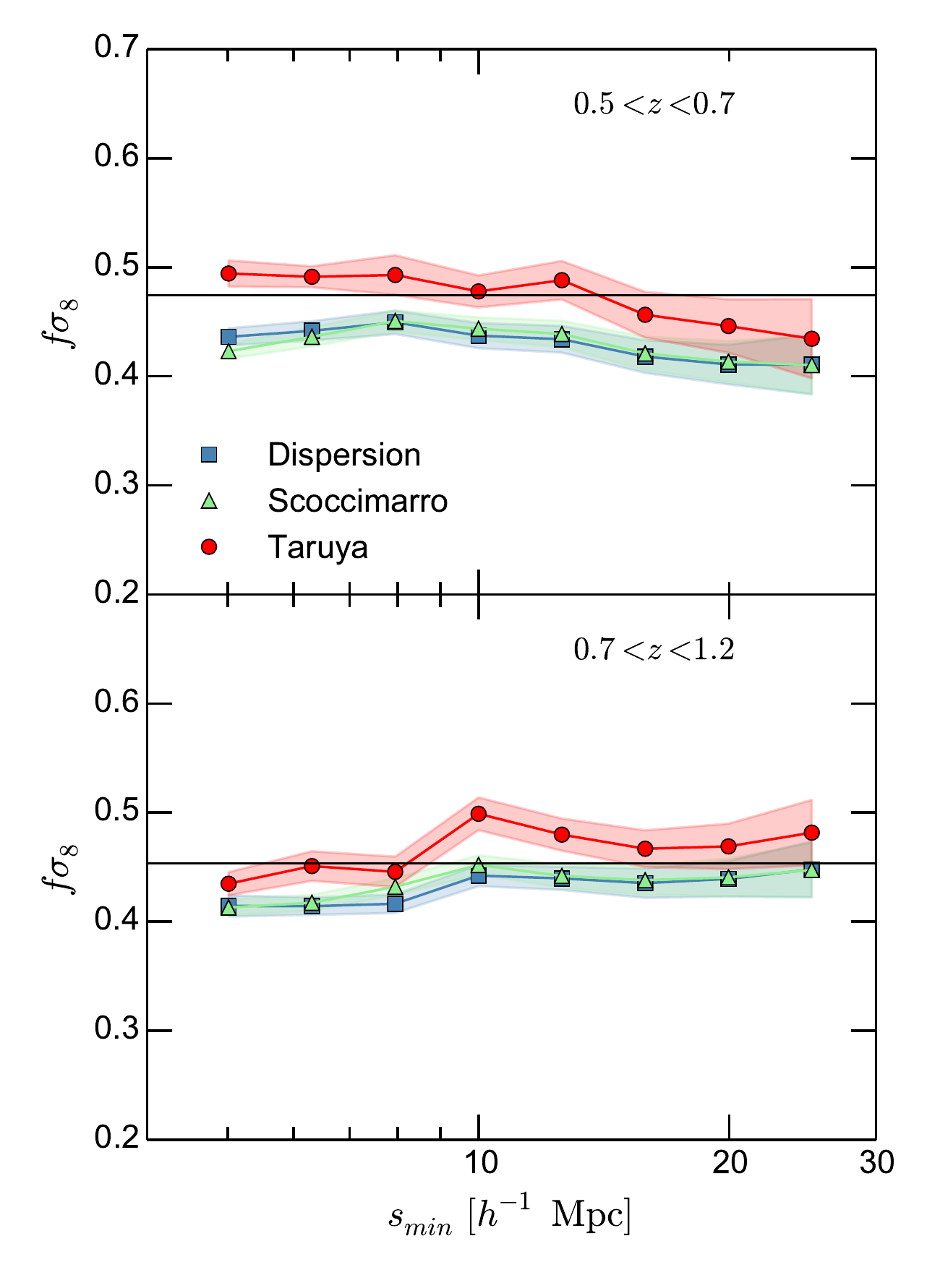}}
\caption{Left panels: the measured values of $f\sigma_8$ from the
  VIPERS survey in the two redshift bins, using the TNS model in its
  optimal set-up that we derived in section~\ref{sec:model_tests}
  (double damping factor: free lorentzian + fixed gaussian), as a
  function of the minimum fitting scale $s_{\rm min}$. The maximum
  fitting scale $s_{\rm max}$ is always fixed at $50\mhmpc$.  The
  shaded area gives the statistical error at each $s_{\rm min}$ for
  the TNS model, as derived from the mocks. Right panels: The same
  measurements performed on the average of the mocks, i.e. plotting
  the results of Fig.~\ref{fig:syst_obs_mocks} but showing explicitly
  the recovered values of $f\sigma_8$, to ease comparison to the data
  results on the left. }
\label{fig:data_smin}
\end{figure*}

We now turn to the case including fully realistic observing
conditions. This means considering the target selection (masks, TSR,
SSR, etc.) and limiting the samples to the same redshift ranges
covered by the data and including redshift errors. The results that we
obtain are shown in Fig. \ref{fig:syst_obs_mocks}. The trends of the
systematic error as a function of $s_{\rm min}$ are less stable than
in the previous case, although the general behaviour and relative
performances of the models are the same.  The variations gives us an
idea of the impact of the selection function on samples this size.
Again, we see some instability in the TNS model (at least in the bin
$0.7<z<1.2$) when the minimum scale of the fit is chosen around
$s_{\rm min}\simeq 8\mhmpc$.  When we look into the MCMC results in
more detail, we see that in this case the Markov chain tends to drift
towards unrealistic values of $\sigma_8$, which are outside of the
prior range defined in Table \ref{tab:parameters}.  This seems to be
related to the difficulty of TNS model to reach a stable fit in the
region where the quadrupole crosses zero.  As soon as we include
smaller scales (or we move away to larger ones), the regular trend is
recovered.  Nevertheless, we confirm the TNS model as the best
performing one, with systematics $\lesssim5\%$ down to the smallest
probed minimum scales.

Overall, the different tests performed on the mock catalogues indicate
that we can safely use the TNS model with the appropriate damping
functions as well as a minimum fitting scale of $s_{\rm min}=5
\mhmpc$. This is the configuration that we adopt for the analysis of
the data.


\section{VIPERS RSD results}

\begin{table}
\centering
\caption{Adopted priors on the sampling parameters.}
\begin{tabular}{c c}
\hline\hline
Parameters & \upstrut\downstrut Uniform prior\\
\hline
$f$ \upstrut & $[0.2,1.8]$ \\
$\sigma_{12}$ & $[0,8]$ \\
$b$ & $[0.5,5]$ \\
$\sigma_8$ & \downstrut $[0.2,0.65]$ \\
\hline\hline
\end{tabular}
\label{tab:priors}
\end{table}

\begin{table}
\centering
\caption{Values of the growth rate and related parameters in the two
  redshift sub-samples, obtained fitting the monopole and quadrupole
  correlation functions over the range $5\mhmpc<s<50\mhmpc$, using the
  TNS model.  Central values and $68\%$ marginalized errors on
  $\sigma_{12}$, $f\sigma_{8}$, and $b\sigma_{8}$ are reported.}
\begin{tabular}{c c c}
\hline\hline
Parameters\upstrut\downstrut & $0.5 \le z \le 0.7$ & $0.7\le z \le 1.2$ \upstrut\\
\hline
$\sigma_{12}$ & $4.996\pm0.855$ & \upstrut $3.542\pm0.784$\\
$f\sigma_8$ & $0.55\pm0.12$ & $0.40\pm0.11$\\
$b\sigma_8$ & $0.73\pm0.03$ & $0.74\pm0.04$ \downstrut\\
\hline\hline
\end{tabular}
\label{tab:parameters}
\end{table}

We present in this section the results of the RSD analysis of the
VIPERS final dataset. We apply the methodology described in the
previous sections to the VIPERS galaxy sample. In the likelihood 
analysis we impose rather broad uniform priors on the sampling 
parameters. These are reported in Table \ref{tab:priors}. Since 
$f$ and $\sigma_8$ are treated as separate parameters in the 
modelling, despite their intrinsic degeneracy, we need to impose 
sensible priors on them. In fact the most sensitive
prior is that on $\sigma_8$, as it is the main parameter entering the
non-linear modelling of RSD. To define a sensible and realistic prior,
while allowing room for deviations from GR, we base our choice on the
Effective Field Theory of dark energy formalism
\citep{Gubitosi2012,Bloomfield2012,Gleyzes2013}, which allows a
description of various kinds of dark energy models and modifications
of gravity to be expressed in a self-consistent framework that
includes the growth rate of structure \citep{Piazza2013,
  Perenon2015}. The latter work shows that the range spanned by
$\sigma_8(z)$ for stable theories can vary significantly, suggesting a
range [$0.2$, $0.65$] as appropriate to account for early- and
late-time dark energy models at the redshifts covered by VIPERS
(for definitions, see \citejap{Perenon2016}).  This excludes some more
extreme modified gravity models, but avoids non-physical degeneracies
that arise in the likelihood for some particular values of $\sigma_8$
outside of this range. This choice is corroborated by our parallel
complementary analysis using the same data by \citet{delatorre16}, in
which the combination of RSD with galaxy-galaxy lensing constrains
directly $\sigma_8(z)$, allowing a broader prior at the outset.

The $f\sigma_8$ measurements that we finally obtain using our 
standard configuraton and previously discussed parameter priors 
are $f\sigma_8(z=0.6)=0.55\pm0.12$ and 
$f\sigma_8(z=0.86)=0.40\pm0.11$. We consider these as our 
reference measurements in this work and discuss their cosmological 
implications in the next section. The measurements and
best-fitting model monopole and quadrupole correlation functions
obtained in the two considered redshift bins are shown in Fig.~\ref{fig:multipoles}. The corresponding best-fit values for 
the derived parameters are reported in Table~\ref{tab:parameters}. 

\begin{figure}
\centering
\includegraphics[width=9cm]{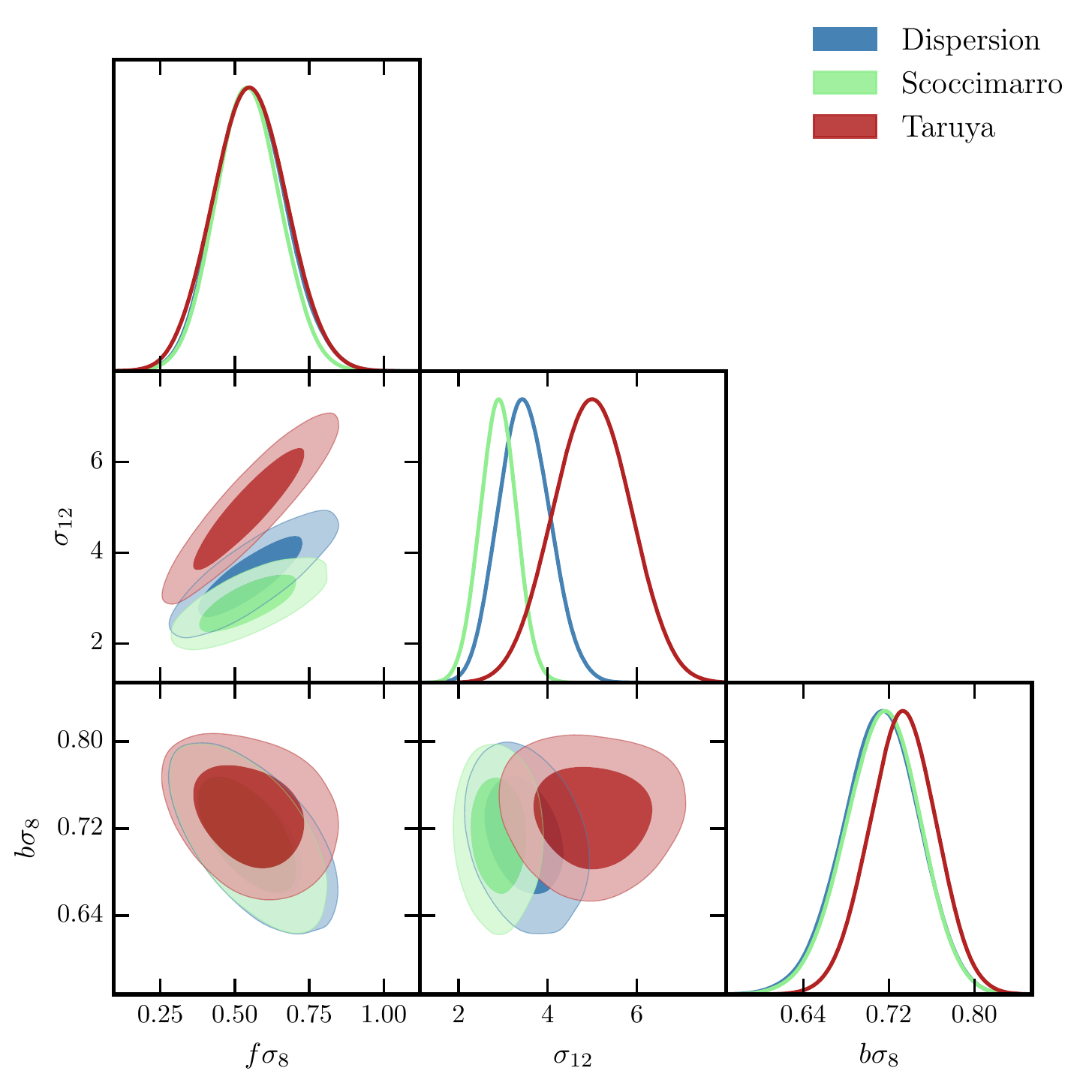}
\caption{The one- and two-dimensional posterior likelihood
  distribution of the derived parameters $f\sigma_8$, $b\sigma_8$ and
  $\sigma_{12}$ for the $0.5 <z < 0.7$ redshift bin. It corresponds to
  the result of the analysis of VIPERS data using dispersion,
  Scoccimarro, and Taruya model and $s_{\rm min}=5$ \hmpc.  The dark-
  and light-shaded areas correspond respectively to the $68\%$ and
  $95\%$ joint two-parameter confidence levels. The lower redshift
  sample shows comparable contours and shapes.}
\label{fig:triangles}
\end{figure}

It is interesting to verify a posteriori whether the trends and
relative RSD model performances as a function of $s_{\rm min}$
established from the mock catalogues are similar to those seen in the
real data.  It is of course clear that any trend will be less
significant, as the data are statistically equivalent to considering
just one of the 153 mock catalogues. In the left panel of
Fig. \ref{fig:data_smin}, we show the result of this exercise, where
the measured values of $f\sigma_8$ as a function of $s_{\rm min}$ are
shown for the different tested models. To ease comparison, we have
reported in the right panel and using the same scale, the
corresponding results from the mock test for the realistic case
(i.e. those of Fig.~\ref{fig:syst_obs_mocks}). Apart from the
different statistical errors, it is surprising to note how the three
tested RSD models provide virtually identical results in the real
data, as opposed to the behaviour seen in the mock catalogues.  Moreover,
it seems that in the data the variation of the $f\sigma_8$
measurements with minimum scale are not driven by the adequacy of the
model down to those scales, but rather by statistical uncertainties in
the measured galaxy correlation functions. The similarity in the
results obtained from the different models is confirmed directly by
the values of the reduced $\chi^2$, which turn out to be very
similar. By directly looking at the posterior likelihood distributions
of the parameters obtained with the three models in
Fig.~\ref{fig:triangles} (for the high-redshift bin), we can see that
each model provides slightly different parameters degeneracies,
although after marginalization, $f\sigma_8$ posterior likelihood
distributions are almost identical for the three RSD models, with only
a slightly larger statistical uncertainty with the TNS model. However,
some trends seen in the mock results are recognised in the data, as
for example the preference of the TNS model in the high-redshift
sample to deliver larger values of $f\sigma_8$ when $s_{\rm min}$ is
close to the zero-crossing scale of the quadrupole.

Finally, it is important to emphasize the global non-linear approach
to RSD that has been used in this analysis. We have used rather small
non-linear scales in the fit, and by adopting a consistent modelling
for the non-linearities in the real-space density and velocity
divergence power spectra, we can obtain further cosmological
insight. The level of non-linearity in our analysis is controlled by
one single parameter, $\sigma_8(z)$, and we find that by letting this
parameter vary, we can partly break the standard degeneracy that
dominates on linear scales between $f$, $\sigma_8$, and $b$
parameters. If we marginalise the posterior likelihood function over the
$\sigma_{12}$, $\sigma_8$, $b$ parameters, we obtain the following
direct growth rate and $\sigma_8$ constraints: 
$[f(z=0.6),\sigma_8(z=0.6)]=[1.048\pm0.199,0.528\pm0.076]$ and 
$[f(z=0.86),\sigma_8(z=0.86)]=[0.742\pm0.179,0.539\pm0.068]$.
A similar approach has been adopted in
\citet{delatorre16}, where this is strengthened by additional constraints from galaxy-galaxy lensing. In particular, the latter allows improving $\sigma_8$ constraints while keeping similar uncertainties on $f$. A detailed discussion of these results is given in \citet{delatorre16}. Overall, these findings demonstrate the additional constraining power encapsulated in quasi-linear scales, which can be used to break degeneracies and further improve the precision of measurement of the growth rate of structure.

\section{Discussion and conclusions}

\begin{figure*}
\centering
\includegraphics[width=13cm]{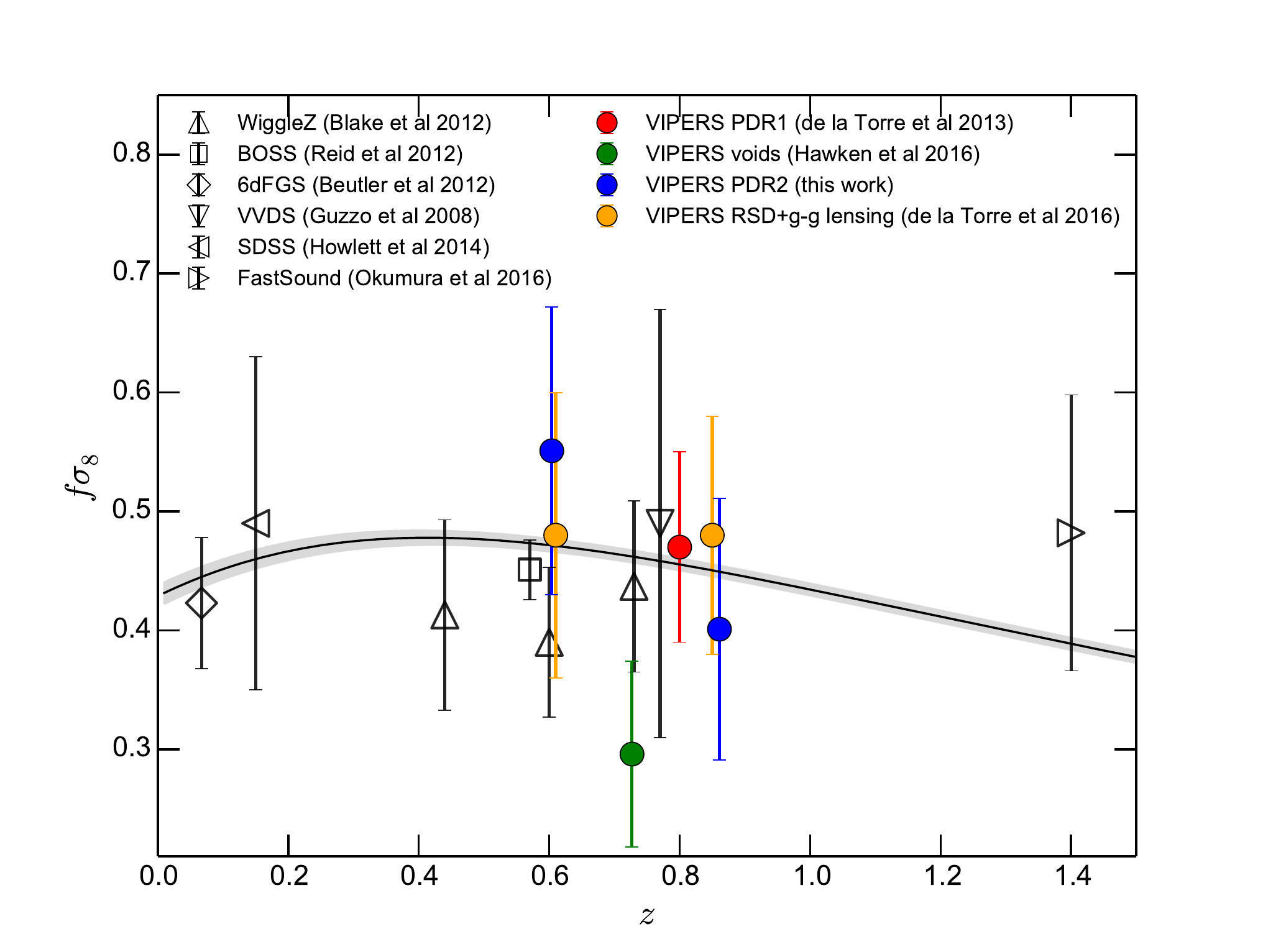}
\caption{Plot of $f\sigma_8$ versus redshift, showing the VIPERS
  results together with a compilation of recent measurements. The
  previous results from 2dFGRS \citep{hawkins03}, 2SLAQ
  \citep{ross07}, VVDS \citep{guzzo08}, SDSS LRG
  \citep{cabre09,samushia12}, WiggleZ \citep{blake11}, BOSS
  \citep{reid12}, 6dFGS \citep{beutler12} and FastSound \citep{okumura16} surveys are shown with
  the different symbols (see inset). The solid curve and associated
  error correspond to the prediction for General Relativity in a
  $\Lambda \rm{CDM}$ model set to Planck 2015 cosmological
  parameters \citep{planck15}.}
\label{fig20}
\end{figure*}

The measurements of the growth rate of structure times $\sigma_8$ that 
we obtained are 
\begin{eqnarray}
f\sigma_8(z=0.6)&=&0.55\pm0.12 \\ 
f\sigma_8(z=0.86)&=&0.40\pm0.11. 
\end{eqnarray}
These values are confronted in Fig.~\ref{fig20} with different 
measurements, including results from other surveys, the VIPERS earlier 
PDR-1 dataset, and parallel works analysing with complementary 
techniques analogous subsets of the VIPERS PDR-2 dataset.  It may look surprising that there is no appreciable improvement in the error bars between the former measurement from the PDR-1 \citep[red circle]{delatorre13} and the new PDR-2 estimate in a comparable redshift bin, despite a $\sim 30\%$ increase in the sample size. As discussed in \citet{delatorre12}, this is essentially a price to pay for the more sophisticated treatment of nonlinear effects through the TNS model, which increases the degrees of freedom. 

The parallel PDR-2 results include measurements obtained from the combination 
of RSD with galaxy-galaxy lensing
\citep{delatorre16} or using the void-galaxy cross-correlation
\citep{hawken16}.  In forthcoming papers, we shall additionally
present further pieces of this combined approach, using specific
colour-selected subsamples \citep{mohammad17} or the linearised
density field in Fourier space \citep{wilson16}, to minimise the need
for non-linear corrections.  All these papers represent complementary
approaches towards understanding the current limitations we face in
our ability to extract in practice the value of these parameters from
the modelling of RSD. 

The values measured by these different techniques on the same VIPERS
data as well as from other surveys at similar redshifts are virtually
all compatible within 1-$\sigma$ and agree with the predictions of a
$\Lambda$CDM model governed by Einstein gravity. But on a larger sample,
with much smaller statistical errors, greater care would be needed to 
test for possible
systematic biases that might still be hidden in one or more of the
analyses.  The application of such a variety of approaches to VIPERS
has been made possible by the specific properties of the survey, in
particular its dense sampling and rich content of information.  With a
sparse sampled survey, which is the approach of most of the cosmologically-oriented surveys, it would have been impossible to characterise accurately 
the density field and apply the clipping linearisation technique of
\citet{wilson16}, or reliably detect cosmic voids such as those used
in \citet{hawken16}.  At the same time, a survey with limited imaging
information would not permit investigation of the selection of optimal
sub-populations (or the combination of different ones), as we are
pursuing in \citet{mohammad17}, or exploit the combination of RSD with
lensing, as we have done in \citet{delatorre16} and which should be
exploited to the fullest by Euclid mission \citep{laureijs11} in the
next decade. We therefore believe that the detailed investigation of the properties of RSD within VIPERS should serve as a valuable foundation for
next-generation studies of greater statistical power.

\begin{acknowledgements}

We acknowledge the crucial contribution of the ESO staff for the
management of service observations. In particular, we are deeply
grateful to M. Hilker for his constant help and support of this
program. Italian participation to VIPERS has been funded by INAF
through PRIN 2008, 2010, 2014 and 2015 programs. LG and BRG
acknowledge support from the European Research Council through grant
n.~291521. OLF acknowledges support from the European Research Council
through grant n.~268107. JAP acknowledges support of the European
Research Council through the COSFORM ERC Advanced Research Grant (\#
670193). GDL acknowledges financial support from the European Research
Council through grant n.~202781. RT acknowledges financial support
from the European Research Council through grant n.~202686. AP, KM,
and JK have been supported by the National Science Centre (grants
UMO-2012/07/B/ST9/04425 and UMO-2013/09/D/ST9/04030). EB, FM and LM
acknowledge the support from grants ASI-INAF I/023/12/0 and PRIN MIUR
2010-2011. LM also acknowledges financial support from PRIN INAF
2012. SDLT and MP acknowledge the support of the OCEVU Labex
(ANR-11-LABX-0060) and the A*MIDEX project (ANR-11-IDEX-0001-02)
funded by the "Investissements d'Avenir" French government program
managed by the ANR. Research conducted within the scope of the HECOLS
International Associated Laboratory, supported in part by the Polish
NCN grant DEC-2013/08/M/ST9/00664. TM and SA acknowledge financial
support from the ANR Spin(e) through the French grant
ANR-13-BS05-0005.

\end{acknowledgements}

\bibliographystyle{aa}
\bibliography{biblio_gg,biblio_VIPERS_v3}


\end{document}